\documentclass[aps,prx,showkeys,reprint]{revtex4-2}

\usepackage[utf8]{inputenc}
\usepackage[T1]{fontenc}
\usepackage{amsmath}
\usepackage{amssymb}
\usepackage{amsfonts}
\usepackage{upgreek}
\usepackage{mathtools}
\usepackage{float}
\usepackage{graphicx}
\usepackage{tabularx}
\usepackage{multirow}
\usepackage[labelfont=bf]{caption}

\usepackage[hidelinks]{hyperref}

\begin{document}

\title{Lifelong Machine Learning Potentials for Chemical Reaction Network Explorations}

\author{Marco Eckhoff}
\email{eckhoffm@ethz.ch}
\affiliation{ETH Zurich, Department of Chemistry and Applied Biosciences, Vladimir-Prelog-Weg 2, 8093 Zurich, Switzerland.}
\author{Markus Reiher}
\email{mreiher@ethz.ch}
\affiliation{ETH Zurich, Department of Chemistry and Applied Biosciences, Vladimir-Prelog-Weg 2, 8093 Zurich, Switzerland.}

\date{September 7, 2025}

\begin{abstract}

Recent developments in computational chemistry facilitate the automated quantum chemical exploration of chemical reaction networks for the in-silico prediction of synthesis pathways, yield, and selectivity. However, the underlying quantum chemical energy calculations require vast computational resources, limiting these explorations severely in practice. Machine learning potentials (MLPs) offer a solution to increase computational efficiency, while retaining the accuracy of reliable first-principles data used for their training. Unfortunately, MLPs will be limited in their generalization ability within chemical (reaction) space, if the underlying training data are not representative for a given application. Within the framework of automated reaction network exploration, where new reactants or reagents composed of any elements from the periodic table can be introduced, this lack of generalizability will be the rule rather than the exception. Here, we therefore evaluate the benefits of the lifelong MLP concept in this context. Lifelong MLPs push their adaptability by efficient continual learning of additional data. We propose an improved learning algorithm for lifelong adaptive data selection yielding efficient integration of new data while previous expertise is preserved. In this way, we can reach chemical accuracy in reaction search trials.

\end{abstract}


\maketitle

\section{Introduction}

In-silico prediction of chemical processes including yield and selectivity can be a key to improving the efficiency of chemical processes and their sustainability \cite{Temkin1996, Fialkowski2005, Szymkuc2016}. However, the reliable representation of reaction kinetics requires knowledge about all possible reactive events. Consequently, large networks of reactions emerge for almost all relevant chemical processes.

The exploration of chemical reaction networks (CRNs) with quantum chemical methods therefore causes immense computational costs in order to identify thousands of stable intermediates and their connecting transition state structures \cite{Feinberg2019, Rappoport2014, Kim2018, Unsleber2020, Baiardi2022, Steiner2022, Ismail2022, Wen2023, Margraf2023, Unsleber2023, Woulfe2024}, which are stationary points on a Born-Oppenheimer potential energy surface. Search trials across this surface can be performed either by explicit construction of potentially reactive complexes \cite{Zimmerman2013, Bergeler2015, Habershon2015, Sameera2016, Simm2017, Dewyer2018, Simm2019, Maeda2021, Unsleber2022} or through molecular dynamics driven searches \cite{Wang2014, Martinez-Nunez2015, Doentgen2015, Vazquez2018, Debnath2019, Grimme2019}. However, both approaches require an enormous number of first-principles single-point calculations, which is a challenge that is also persistent in other tasks of computational chemistry, biology, and materials science such as virtual high-throughput screening \cite{Shoichet2004, Pyzer-Knapp2015}.

Accordingly, fast quantum chemical approaches are needed. However, they are plagued by drastic approximations that compromise not only energies but also molecular structures \cite{Christensen2016, Dral2016}. As a consequence, a CRN constructed with an approximate quantum chemical approach (such as tight-binding density functional or semi-empirical theories) does not necessarily represent the network that would be obtained with a more reliable approach (such as density functional theory (DFT)). Hence, node and connection fidelity can be compromised. Local refinement by applying more accurate methods for the a-posteriori reoptimization of stable intermediates and transition states could improve molecular structures and, most importantly, their relative energies \cite{Proppe2016, Simm2018, Bensberg2024}. However, this requires a CRN to be qualitatively correct. Hence, the stationary points optimized with a fast approach must be complete as must be their connections by elementary reaction steps. It is therefore key to devise first-principles-based methods that are (I) fast to allow for efficient reactive event screening and that have (II) high structural fidelity so that the topology and structure of a CRN is not compromised. Hence, machine learning potentials (MLPs) hold the greatest promise to yield both, efficiency and structure fidelity \cite{Behler2016, Bartok2017, Deringer2019, Noe2020, Westermayr2021a, Kaeser2023, Yang2024, Kuryla2025}.

However, the requirements for MLPs in CRN explorations are challenging. (I) They must be quickly initializable and (II) they must not come with an unbearable overhead for their training. It is precisely the goal of universal MLPs to deliver a parametrization that is as general as possible to overcome the initialization issue and to avoid training at runtime \cite{Chen2022, Batatia2023, Kovacs2023, Takamoto2023, Choudhary2023, Merchant2023, Deng2023, Zhang2024, Yang2024a, Anstine2025, Wood2025}. To achieve this goal, models with universal structure representations are trained on very large and diverse reference data sets. Unfortunately, an out-of-the-box approach without further fine-tuning of the universal foundation model on some specialized data of the given chemical system can suffer from insufficient accuracy \cite{Focassio2024, Yu2024}. In order to tackle this challenge, we therefore proposed the concept of lifelong machine learning potentials (lMLP) to alleviate the overhead of training \cite{Eckhoff2023}. In contrast to conventionally trained MLPs, lMLPs can continuously adapt to additional data without forgetting previous knowledge, while they do not require training again on all previous data. Hence, they can offer high accuracy for the parts of chemical space studied, but they may still require additional quantum chemical calculations during their application. We note that the lifelong learning process cannot only start from scratch but also from the parametrization of a pre-trained universal foundation model. In contrast to conventional fine-tuning of such a foundation model, lifelong learning can be continued for much more than one fine-tuning iteration.

In this work, we assess the reliability, accuracy, and efficiency of lMLPs for CRN explorations. Section \ref{sec:Methods} summarizes the concept of lMLPs and introduces an improved algorithm for lifelong adaptive data selection. The computational details are compiled in Section \ref{sec:Computational_Details}. Section \ref{sec:Results_and_Discussion} starts with a presentation of the $\text{HCN}+\text{H}_2\text{O}$ CRN studied in this work. Afterwards, continual learning of an lMLP on the CRN data is analyzed and compared to conventional iterative learning and DFT reference data. 

\section{Methods}\label{sec:Methods}

\subsection{Lifelong Machine Learning Potentials}

The concept of lMLPs \cite{Eckhoff2023} introduces continual or lifelong machine learning \cite{Chen2018, Parisi2019} into the MLP training process. An efficient online learning process can greatly advance the adaptability of MLPs to yield high accuracy and general applicability for every chemical structure within a reasonable period of time. Consequently, lMLPs may require some reference calculations on-the-fly during their application in simulations, but these additional training data can be integrated with low cost. However, continual machine learning is a challenge in itself \cite{Grossberg2013, Goodfellow2015}, especially in the single incremental task scenario \cite{Maltoni2019} which is given for the prediction of the potential energy surface for more and more chemical structures. We note that an alternative approach for on-the-fly learning of MLPs can be obtained by Bayesian inference \cite{Jinnouchi2019, Jinnouchi2019a}.

\begin{figure}[htb!]
\centering
\includegraphics[width=\columnwidth]{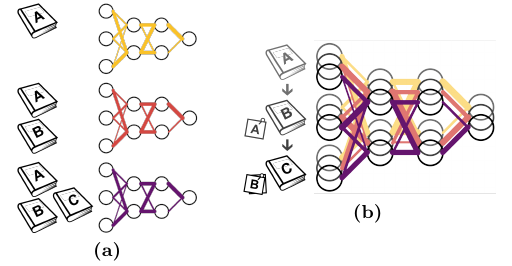}
\caption{\textbf{(a)} In iterative learning, a new machine learning model is trained from scratch on all data to integrate additional training data. \textbf{(b)} In lifelong learning, the training of a model is continued employing added data and only a subset of previous data to prevent forgetting. In this way, previously acquired knowledge of the model is exploited. We note that lifelong learning is here only based on rehearsal of data, while it can also exploit model parameter regularization and the model architecture.}\label{fig:iterative_and_lifelong_learning}
\end{figure}

In conventional iterative learning, each extension of the reference data requires training from scratch on all data to obtain an improved machine learning model. For example, to obtain a model that cannot only represent data A but also data B, a new model (red model in Figure \ref{fig:iterative_and_lifelong_learning}) is trained on both data sets simultaneously (Figure \ref{fig:iterative_and_lifelong_learning} (a)). Further extension to data C (purple model in Figure \ref{fig:iterative_and_lifelong_learning}) then requires training on all data A+B+C. Consequently, the effort for model adaptions by the same amount of added data increases more and more for larger training data sets. If the training of the previous model was continued only on the added data, so-called catastrophic forgetting would occur, i.e., the previous knowledge vanishes due to the optimization of the model parameters only on new data.

Therefore, continual learning introduces training strategies to mitigate catastrophic forgetting by rehearsal of essential training data, regularization of model parameters, and/or the model architecture. For example, training of a model based on data A (yellow model in Figure \ref{fig:iterative_and_lifelong_learning}) can be continued by learning the added data B and a small subset of the previous data A (Figure \ref{fig:iterative_and_lifelong_learning} (b)). This subset only needs to ensure that the already existing knowledge will be retained, i.e., previously acquired knowledge will be exploited and the training cost is reduced. It can be chosen during the previous training process applying, for example, lifelong adaptive data selection (lADS) \cite{Eckhoff2023} (see Section 2.2). Subsequently, the learning process can be further continued by training on added data C and a subset of A+B. In this incremental learning approach, training on additional data can be efficient even for large data sets because training again on the entire data set A+B+C can be avoided. Moreover, several continual learning strategies can be combined. For example, the model parameter regularization of the continual resilient (CoRe) optimizer \cite{Eckhoff2023, Eckhoff2024} can be applied in addition to lADS.

\begin{figure}[htb!]
\centering
\includegraphics[width=0.8\columnwidth]{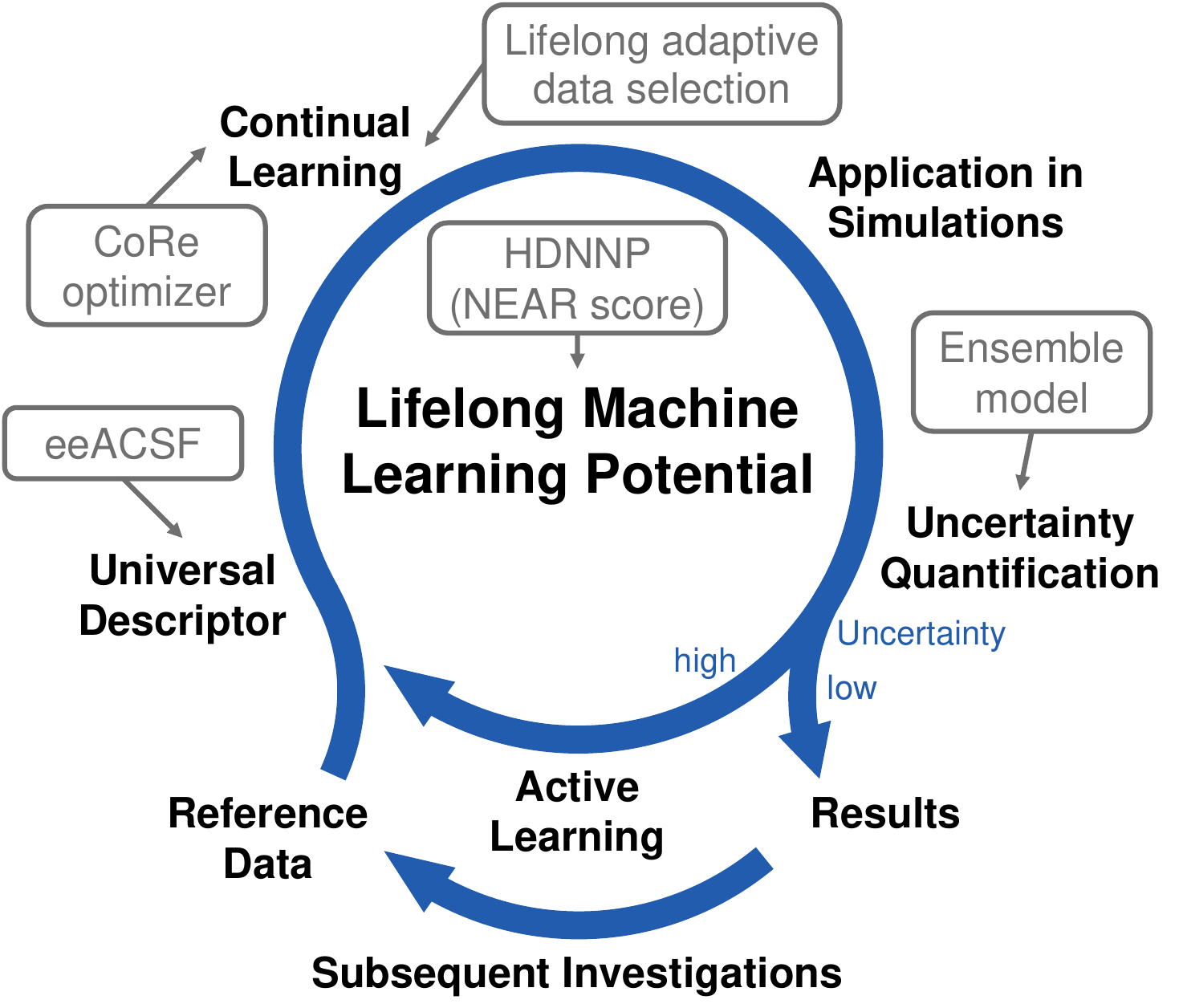}
\caption{Key conceptual elements of lMLPs denoted in black font. Example methods are denoted in gray (see main text for details). Reference data need to be represented by a universal descriptor to be learned by the lMLP. If simulations encounter lMLP predictions of high uncertainty, active learning tackles this issue in the framework of a continual learning approach for an efficient adaption of the lMLP. 
Moreover, results obtained in such simulations may point to follow-up simulations, e.g., on related systems, which may then require training on additional data coming in. Data points integrated by continual learning neither require training from scratch, nor on all data.
}\label{fig:lMLP_concept}
\end{figure}

With the help of continual learning, a closed and efficient loop in the evolution and application of an lMLP can be obtained (Figure \ref{fig:lMLP_concept}). The process starts with some initial reference data. These data can be represented through universal descriptors such as element-embracing atom-centered symmetry functions (eeACSFs) \cite{Eckhoff2023}, so that they can be learned by an lMLP utilizing, for example, the high-dimensional neural network potential (HDNNP) method \cite{Behler2007, Behler2017, Behler2021} (see Supporting Information Section S1.1 of this work for an overview). Network expressivity by activation rank (NEAR) \cite{Husistein2024} can help with the choice of the neural network architecture. We note that the lMLP concept can also be employed with other methods, including approaches which combine descriptor and potential into a single learnable representation. Moreover, the initial parametrization of the MLP can be taken from a pre-trained universal foundation model instead of employing some random weight initialization. In any case, the representation must be able to handle any chemical structure to yield a generally applicable lMLP. For example, many MLP descriptors cannot efficiently represent a system with many different chemical elements because their descriptor vector size increases with an increasing number of different elements. As a consequence, the application of conventional atom-centered symmetry functions \cite{Behler2011} is often restricted to systems with at most four different elements because of computational costs. By combination of (molecular) structure information with element information of the periodic table, eeACSFs can overcome this limitation \cite{Gastegger2018, Eckhoff2023} (see Supporting Information Section S1.2 of this work for an overview). Moreover, in Supporting Information Section S1.2 of this work we propose alternative functions based on the bump function to represent the radial and angular structure within eeACSFs and we introduce the cube root-scaled-shifted (crss) scaling function for eeACSFs.

To be able to deal with molecules with different total charges or spin multiplicities, individual lMLPs for each state can be trained. Alternatively, the second-generation HDNNP base model can be replaced by other models that can handle multiple states (and long-range interactions beyond the cutoff radius employed in the MLP descriptor), such as fourth-generation HDNNPs \cite{Ko2021}. Moreover, continual learning can also be applied to train atomic-property machine learning models (e.g., for atomic charges and spins) to calculate these efficiently, e.g., during molecular dynamics simulations.

Uncertainty quantification can be exploited in applications of the trained model to probe a pre-defined accuracy on the fly and to identify the need for new data points. In fact, uncertainty quantification is necessary to enable application of an lMLP at every training stage because it provides the confidence interval for analysis of the results. Uncertainties can be obtained, for example, by an ensemble or committee model \cite{Peterson2017, Smith2018, Musil2019, Imbalzano2021}. This can provide a quantitative estimate of small errors, while large errors can only be flagged. However, an indication of large errors is sufficient because only the need for additional data must be revealed. Those are then produced and fed into the continual training process until all errors are found to be sufficiently small.

In case of high uncertainties, active learning with a query-by-committee approach can be applied to complete the training data sampling. In active learning of MLPs, (I) missing training data are identified during MLP application based on the uncertainty assessed, (II) uncertainly predicted chemical structures are recalculated by the reference method, and (III) the MLP is retrained. Continual learning can be applied to speed up the retraining process compared to the conventional application of iterative learning during active learning. Hence, lifelong learning and active learning therefore complement one another. In addition, chemical insights resulting from the simulations may lead to subsequent tasks, which require reference data of further chemical systems or reactions. These can be generated by various approaches in order to extend the model (such as (random) variations of experimental or handcrafted structures or ab initio molecular dynamics simulations). Also here, continual learning can fasten the adaption of the model to these data.

Therefore, we note that lifelong or continual learning can be a sub-task of active learning, but generally speaking it refers to a continual model (re-)training process and can be applied without training data generation by active learning. Continual learning itself does not include a training data generation workflow in contrast to active learning. It allows us to avoid inefficient model training on all previous training data from scratch again.

\subsection{Lifelong Adaptive Data Selection}

Lifelong adaptive data selection (lADS) is a continual learning algorithm utilizing rehearsal of previous training data. Its goal is a continuous reduction of the training data to distill important data for (re-)training in continual learning. Moreover, it includes a mechanism to remove inconsistent data, which is necessary in online learning due to limited options for data pre-processing. In addition, lADS ranks the data points according to their importance for training to improve learning efficiency.

\begin{figure}[htb!]
\centering
\includegraphics[width=0.8\columnwidth]{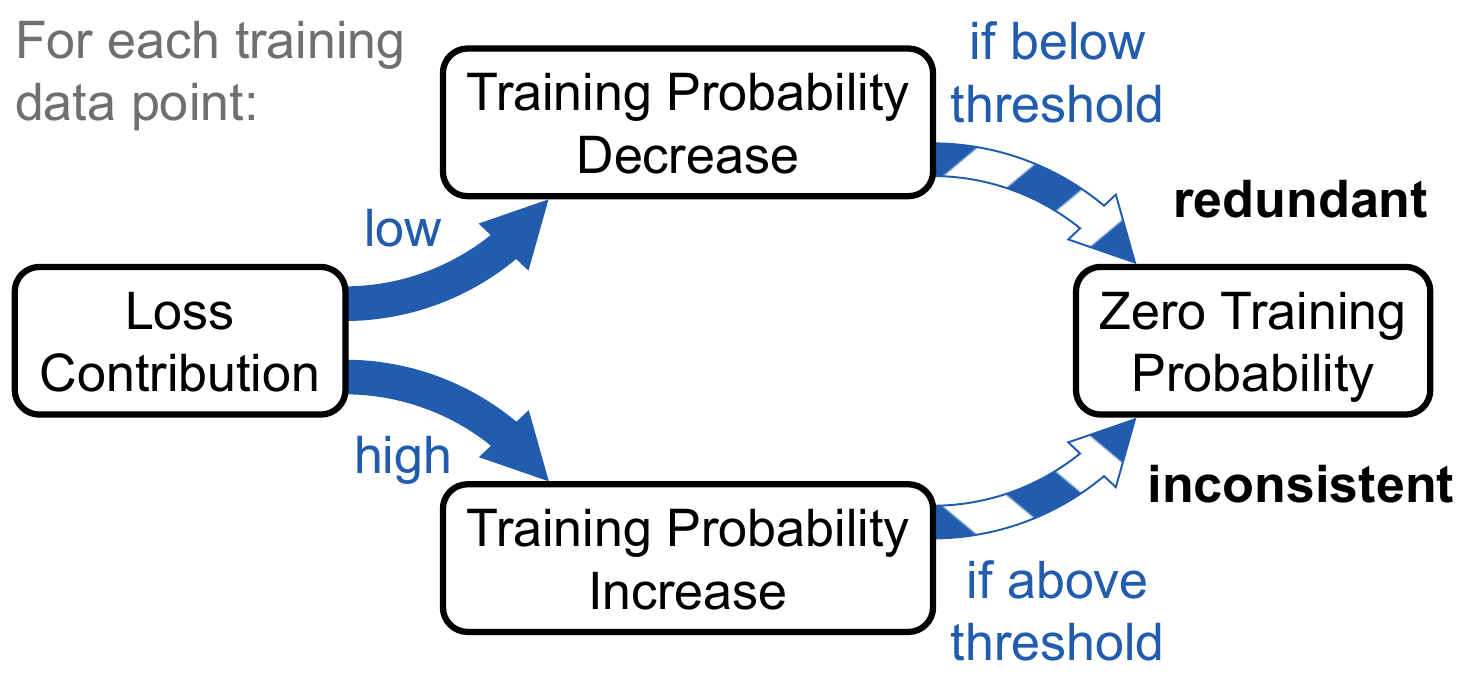}
\caption{Simplified scheme of the core concept behind lADS to continuously reduce and clean the data during training. (I) Data is redundant if it is seldom trained but still well represented. (II) Data is likely to be inconsistent if it is very often trained but still badly represented.}\label{fig:lADS_concept}
\end{figure}

The main ideas behind lADS (Figure \ref{fig:lADS_concept}) are that, on the one hand, training data will be redundant if they are seldom trained in continual learning but still well represented. On the other hand, it is likely that data will be inconsistent with the majority of data if they are very often trained but still poorly represented. Consequently, lADS requires only a fraction of all training structures in each training step, applying a biased random selection. The probability for training of a given structure thereby needs to be adjusted according to how well the structure is represented by the model and how stable the quality of this representation is. In this way, both aforementioned ideas can be exploited. Moreover, learning can also focus on the insufficiently represented training data. 

To determine whether a structure is well or poorly represented, the loss is utilized. The loss is the property that is minimized during training, representing the difference between model predictions and reference values. The total loss is a sum of the contributions of all $N_\mathrm{fit}$ structures $r$,
\begin{align}
L_\mathrm{total}=\sum_{r=1}^{N_\mathrm{fit}}\dfrac{q}{N_\mathrm{fit}}\mathcal{L}\left(\dfrac{E^r}{N_\mathrm{atom}^r}\right)+\dfrac{\mathcal{L}\left(\mathbf{F}^r\right)}{3N_\mathrm{atom}^\mathrm{fit,sum}}\ ,\label{eq:total_loss}
\end{align}
with the loss functions $\mathcal{L}$ of the energy $E^r$ and the Cartesian components of the atomic forces $F_{\alpha,n}^r$ of the $N_\mathrm{atom}^r$ atoms $n$ of each structure $r$,
\begin{align}
\mathcal{L}\left(\dfrac{E^r}{N_\mathrm{atom}^r}\right)=\left(\dfrac{E^r-E^{\mathrm{ref},r}}{N_\mathrm{atom}^r}\right)^2\ ,\\
\mathcal{L}\left(\mathbf{F}^r\right)=\sum_{n=1}^{N_\mathrm{atom}^r}\sum_{\alpha=1}^3\left(F_{\alpha,n}^r-F_{\alpha,n}^{\mathrm{ref},r}\right)^2\ .
\end{align}
We apply here the squared deviation between predicted and reference values. The loss of energy and forces can be balanced by the hyperparameter $q$. Consequently, the representation quality can also be assessed by a loss vector $\mathbf{L}$ containing the contributions of each structure separately. To address the representation quality, we further split the loss contributions of energy and forces in the vectors $\mathbf{L}^E$ and $\mathbf{L}^F$, whereby, compared to the total loss $L_\mathrm{total}$, the energy contributions are not divided by $N_\mathrm{fit}$ and the force contributions are divided by $N_\mathrm{atom}^r$ instead of $N_\mathrm{atom}^\mathrm{fit,sum}$. The latter is the total number of all atoms in all structures to be fitted. 

The main difference of the lADS algorithm presented in this work compared to its original implementation in Reference \cite{Eckhoff2023} is (I) the separate consideration of energy and force loss contributions. (II) the advanced algorithm adjusts the number of fitted structures per training step. (III) it employs an improved scheme to calculate the training probabilities and (IV) to update the selection determining properties. (V) it includes a maximum number of structures that can be classified as redundant per step and (VI) adds a scheme for fast integration of additional training data. In addition, (VII) it enables removing loss gradient contributions of inconsistent training data in the history of the CoRe optimizer. The following four paragraphs will explain all details of the lADS algorithm.

\subsubsection*{Choice of Data to be Fitted}

\begin{table}[htb!]
\centering
\begin{tabular}{p{8.25cm}}
\hline\vspace{-0.125cm}
\textbf{Algorithm 1:} Selection of a training data subsample which will be fitted in the training epoch. We note that in all algorithm notations the vector operations are element-wise and assignments including conditions affect only vector entries which fulfill the conditions.\\
We use the following naming conventions: $\mathbf{D}$ is a vector of data points. $\mathbf{L}$ are loss contributions for each data point. $\mathbf{P}$ are probabilities for each data point to be selected for training. $\mathbf{S}_\mathrm{hist}$ are the adaptive selection factors employed in the probability determination of each data point. $N$ is a count of a quantity. Hyperparameters are defined in the main text.\vspace{0.125cm}\\
\hline\vspace{-0.125cm}
$N_\mathrm{fit}\leftarrow\mathrm{max}\left[1,\ \lfloor p_\mathrm{fit}\cdot\mathrm{dim}\left(\mathbf{S}_\mathrm{hist}^{>0,E}\right)\rfloor\right]$\vspace{0.075cm}\\
$N_\mathrm{good}\leftarrow\lfloor p_\mathrm{good}\cdot N_\mathrm{fit}\rfloor$\vspace{0.075cm}\\
$N_\mathrm{bad}\leftarrow N_\mathrm{fit}-N_\mathrm{good}$\vspace{0.075cm}\\
$L_\mathrm{old}^{\mathrm{max},E/F}\leftarrow\mathrm{max}\left(\mathbf{L}_\mathrm{old}^{\backslash\mathrm{NaN},S_\mathrm{hist}^{r,E}>0,E/F}\right)$\vspace{0.075cm}\\
$\mathrm{if}\ L_\mathrm{old}^{\mathrm{max},E/F}>0$\vspace{0.075cm}\\
$\quad\quad\mathbf{P}_\mathrm{loss}^{E/F}\leftarrow\dfrac{\mathbf{L}_\mathrm{old}^{E/F}}{L_\mathrm{old}^{\mathrm{max},E/F}}$\vspace{0.075cm}\\
$\quad\quad\mathbf{P}_\mathrm{loss}^{E/F}\leftarrow\dfrac{\overline{L}_\mathrm{old}^{E/F}}{L_\mathrm{old}^{\mathrm{max},E/F}}\ \ \mathrm{if}\ P_\mathrm{loss}^{r,E/F}<\dfrac{\overline{L}_\mathrm{old}^{E/F}}{L_\mathrm{old}^{\mathrm{max},E/F}}$\vspace{0.075cm}\\
$\mathrm{else}$\vspace{0.075cm}\\
$\quad\quad\mathbf{P}_\mathrm{loss}^{E/F}\leftarrow\mathbf{1}$\vspace{0.075cm}\\
$\mathbf{P}_\mathrm{loss}\leftarrow\left(\mathbf{P}_\mathrm{loss}^E\cdot\mathbf{P}_\mathrm{loss}^F\right)^{\tfrac{1}{2}}$\vspace{0.075cm}\\
$\mathbf{P}_\mathrm{bad}\leftarrow\mathbf{S}_\mathrm{hist}^E\cdot\left|\mathbf{S}_\mathrm{hist}^F\right|\cdot\mathbf{P}_\mathrm{loss}$\vspace{0.075cm}\\
$P_\mathrm{bad}^\mathrm{max}\leftarrow\mathrm{max}\left[S_\mathrm{hist}^\mathrm{max},\ \mathrm{max}\left(\mathbf{P}_\mathrm{bad}^{\backslash\mathrm{NaN}}\right)\right]$\vspace{0.075cm}\\
$\mathbf{P}_\mathrm{bad}\leftarrow P_\mathrm{bad}^\mathrm{max}\ \ \mathrm{if}\ P_\mathrm{bad}^r=\mathrm{NaN}$\vspace{0.075cm}\\
$\mathbf{P}_\mathrm{bad}\leftarrow 0\ \ \mathrm{if}\ S_\mathrm{hist}^{r,E}\leq0$\vspace{0.075cm}\\
$\mathbf{P}_\mathrm{bad}\leftarrow\dfrac{\mathbf{P}_\mathrm{bad}}{\mathrm{sum}\left(\mathbf{P}_\mathrm{bad}\right)}$\vspace{0.075cm}\\
$\mathbf{D}_\mathrm{fit}\leftarrow\mathrm{random{\_}choice}\left(\mathbf{D},\ \mathbf{P}_\mathrm{bad},\ N_\mathrm{bad}\right)$\vspace{0.075cm}\\
$\mathbf{P}_\mathrm{good}\leftarrow\mathbf{S}_\mathrm{hist}^{\backslash\mathrm{fit},E}\cdot\left|\mathbf{S}_\mathrm{hist}^{\backslash\mathrm{fit},F}\right|\cdot\left(1-\mathbf{P}_\mathrm{loss}^{\backslash\mathrm{fit}}\right)$\vspace{0.075cm}\\
$P_\mathrm{good}^\mathrm{min}\leftarrow\mathrm{min}\left[S_\mathrm{hist}^\mathrm{min},\ \mathrm{min}\left(\mathbf{P}_\mathrm{good}^{\backslash\mathrm{NaN},>0}\right)\right]$\vspace{0.075cm}\\
$\mathbf{P}_\mathrm{good}\leftarrow P_\mathrm{good}^\mathrm{min}\ \ \mathrm{if}\ P_\mathrm{good}^r=\mathrm{NaN}\lor P_\mathrm{loss}^{\backslash\mathrm{fit},r}=1$\vspace{0.075cm}\\
$\mathbf{P}_\mathrm{good}\leftarrow 0\ \ \mathrm{if}\ S_\mathrm{hist}^{\backslash\mathrm{fit},r,E}\leq0$\vspace{0.075cm}\\
$\mathbf{P}_\mathrm{good}\leftarrow\dfrac{\mathbf{P}_\mathrm{good}}{\mathrm{sum}\left(\mathbf{P}_\mathrm{good}\right)}$\vspace{0.075cm}\\
$\mathbf{D}_\mathrm{fit}\leftarrow\mathbf{D}_\mathrm{fit}\cup\mathrm{random{\_}choice}\left(\mathbf{D}^{\backslash\mathrm{fit}},\ \mathbf{P}_\mathrm{good},\ N_\mathrm{good}\right)$\vspace{0.125cm}\\
\hline
\end{tabular}
\end{table}

The selection of the training structures to be fitted is given in Algorithm 1 and will be explained step-by-step in this paragraph. In each training step, the lADS algorithm utilizes a fraction of $p_\mathrm{fit}$ of all training structures that are still available for training, i.e., structures $r$ with an adaptive selection factor $S_\mathrm{hist}^{r,E}$ larger than zero. In general, these adaptive selection factors determine the training probability. They incorporate the representation quality history for the energy and forces of each structure. The starting value for each structure is $S_\mathrm{hist}^{r,E/F}=1$. From the $N_\mathrm{fit}$ structures, a fraction of $p_\mathrm{good}$ is selected from well represented structures, while the remaining $N_\mathrm{bad}$ structures are not yet well represented. $p_\mathrm{good}$ is initialized before the first step as zero and will be adapted as shown in Algorithm 2 in the next paragraph. By employing good and bad data, the stability--plasticity balance of retaining old expertise and integrating new knowledge can be improved. Moreover, training on good data is required for lADS to sort out redundant, well represented data.

To obtain the training probabilities, the maximum loss contributions for energies and forces $L_\mathrm{old}^{\mathrm{max},E/F}$ are determined among all previously calculated loss contributions of structures available for training. We note that the loss vectors $\mathbf{L}_\mathrm{old}^{E/F}$ are initialized before the first step as $\mathbf{NaN}$, i.e., a vector of not a number (NaN) entries. If there is at least one structure with a loss contribution greater than zero, the probability based on the loss $\mathbf{P}_\mathrm{loss}^{E/F}$ is calculated as the loss vector divided by the maximum loss contribution. Thereby, the minimum value is set to the mean loss contribution $\overline{L}_\mathrm{old}^{E/F}$ divided by the maximum loss contribution. $\mathbf{P}_\mathrm{loss}^{E/F}$ increases the probability that badly represented structures are selected, while providing equal probabilities for well represented structures so that even very well represented structures have a chance of being selected. If no loss contribution is greater than zero, $\mathbf{P}_\mathrm{loss}^{E/F}=\mathbf{1}$ is applied. The probability vectors for energy and forces are multiplied and their square root is taken, resulting in $\mathbf{P}_\mathrm{loss}$. Subsequently, $\mathbf{P}_\mathrm{loss}$ is multiplied by the adaptive selection factor vectors of energies and forces. While $\mathbf{P}_\mathrm{loss}$ contains information about the current representation quality, the adaptive selections factors account for the history of the representation quality. The maximum value of the resulting vector $\mathbf{P}_\mathrm{bad}$ is determined to be $P_\mathrm{bad}^\mathrm{max}$ that has a lower bound of $S_\mathrm{hist}^\mathrm{max}$. The latter also defines the maximum value in $\mathbf{S}_\mathrm{hist}$. All $\mathrm{NaN}$ in $\mathbf{P}_\mathrm{bad}$ are set to $P_\mathrm{bad}^\mathrm{max}$, i.e., structures trained for the first time have the highest probability in $P_\mathrm{bad}$. For unavailable training structures, the respective vector entries are set to zero. The final probability is obtained by dividing the vector by the sum of its entries. Afterwards, $N_\mathrm{bad}$ structures are randomly selected from the training data $\mathbf{D}$, with the respective probabilities $\mathbf{P}_\mathrm{bad}$.

To choose the $N_\mathrm{good}$ structures, $1-\mathbf{P}_\mathrm{loss}$ is multiplied by the adaptive selection factor vectors for those structures that were not yet selected, resulting in $\mathbf{P}_\mathrm{good}$. Therefore, high loss contributions lead to a lower probability in this case. The minimum value of all values larger than 0 is determined as $P_\mathrm{good}^\mathrm{min}$ that has an upper bound of $S_\mathrm{hist}^\mathrm{min}$. The latter also defines the minimum value of the adaptive selection factor of the structures available for training. $P_\mathrm{good}^\mathrm{min}$ will be employed in $\mathbf{P}_\mathrm{good}$ if the vector entry is $\mathrm{NaN}$ or the respective $P_\mathrm{loss}^r$ is $1$, i.e., structures trained for the first time and those with the highest loss have the lowest probability in $P_\mathrm{good}$. Analogously to $\mathbf{P}_\mathrm{bad}$, $\mathbf{P}_\mathrm{good}$ is set to zero for unavailable structures, and it is divided by the sum of its entries. Finally, the resulting random training data choice using the probabilities $\mathbf{P}_\mathrm{good}$ is combined with the data selected using $\mathbf{P}_\mathrm{bad}$. This combined data subsample $\mathbf{D}_\mathrm{fit}$ will be fitted in the training epoch.

\subsubsection*{Update of Selection Properties}

\begin{table}[htb!]
\centering
\begin{tabular}{p{8.25cm}}
\hline\vspace{-0.175cm}
\textbf{Algorithm 2:} Update of the selection-determining properties.\vspace{0.125cm}\\
\hline\vspace{-0.125cm}
$\mathbf{L}_\mathrm{new}^E\leftarrow q\cdot\mathcal{L}\left(\dfrac{\mathbf{E}^\mathrm{fit}}{\mathbf{N}_\mathrm{atom}^\mathrm{fit}}\right)$\vspace{0.075cm}\\
$\mathbf{L}_\mathrm{new}^F\leftarrow\dfrac{\mathcal{L\left(\mathbf{F}^\mathrm{fit}\right)}}{3\mathbf{N}_\mathrm{atom}^\mathrm{fit}}$\vspace{0.075cm}\\
$\mathrm{for}\ i\ \mathrm{in}\ \{1,2,3,4\}$\vspace{0.075cm}\\
$\quad\quad\mathbf{L}_{Ti}^E\leftarrow q\cdot\mathcal{L}\left(\dfrac{T_i\cdot\mathbf{E}^\mathrm{fit}}{\mathbf{N}_\mathrm{atom}^\mathrm{fit}}\right)$\vspace{0.075cm}\\
$\quad\quad\mathbf{L}_{Ti}^F\leftarrow\dfrac{\mathcal{L}\left(T_i\cdot\mathbf{F^\mathrm{fit}}\right)}{3\mathbf{N}_\mathrm{atom}^\mathrm{fit}}$\vspace{0.075cm}\\
$\quad\quad\overline{L}_{Ti}^{E/F}\leftarrow\mathrm{mean}\left(\mathbf{L}_{Ti}^{\backslash\mathrm{NaN},S_\mathrm{hist}^{r,E}\geq-1,E/F}\right)$\vspace{0.075cm}\\
$\mathbf{X}\leftarrow X^r+1\ \ \mathrm{if}\ L_\mathrm{new}^{r,E}>\overline{L}_{T4}^E\lor L_\mathrm{new}^{r,F}>\overline{L}_{T4}^F$\vspace{0.075cm}\\
$\mathbf{X}\leftarrow 0\ \ \mathrm{if}\ L_\mathrm{new}^{r,E}\leq \overline{L}_{T4}^E\land L_\mathrm{new}^{r,F}\leq \overline{L}_{T4}^F$\vspace{0.075cm}\\
$\mathbf{S}_\mathrm{hist}^{E/F}\leftarrow\mathrm{max}\left(1,\ S_\mathrm{hist}^{r,E/F}\right)\ \ \mathrm{if}\ L_\mathrm{new}^{r,E/F}\geq \overline{L}_{T1}^{E/F}$\vspace{0.075cm}\\
$\mathbf{S}_\mathrm{hist}^{E/F}\leftarrow\mathrm{min}\left(S_\mathrm{hist}^{r,E/F},\ 1\right)\ \ \mathrm{if}\ L_\mathrm{new}^{r,E/F}\leq \overline{L}_{T2}^{E/F}$\vspace{0.075cm}\\
$\mathbf{S}_\mathrm{hist}^{E/F}\leftarrow S_\mathrm{hist}^{r,E/F}\cdot F_{--}\ \ \mathrm{if}\ L_\mathrm{new}^{r,E/F}<\overline{L}_{T1}^{E/F}\land L_\mathrm{new}^{r,E/F}\leq L_\mathrm{old}^{r,E/F}$\vspace{0.075cm}\\
$\mathbf{S}_\mathrm{hist}^{E/F}\leftarrow S_\mathrm{hist}^{r,E/F}\cdot F_{-}\ \ \mathrm{if}\ L_\mathrm{new}^{r,E/F}<\overline{L}_{T1}^{E/F}\land L_\mathrm{new}^{r,E/F}>L_\mathrm{old}^{r,E/F}$\vspace{0.075cm}\\
$\mathbf{S}_\mathrm{hist}^{E/F}\leftarrow S_\mathrm{hist}^{r,E/F}\cdot F_+\ \ \mathrm{if}\ \overline{L}_{T2}^{E/F}<L_\mathrm{new}^{r,E/F}\leq \overline{L}_{T3}^{E/F}$\\
$\quad\quad\quad\ \ \ \land L_\mathrm{new}^{r,E/F}>L_\mathrm{old}^{r,E/F}$\vspace{0.075cm}\\
$\mathbf{S}_\mathrm{hist}^{E/F}\leftarrow S_\mathrm{hist}^{r,E/F}\cdot F_{+}\ \ \mathrm{if}\ L_\mathrm{new}^{r,E/F}>\overline{L}_{T3}^{E/F}\land L_\mathrm{new}^{r,E/F}\leq L_\mathrm{old}^{r,E/F}$\vspace{0.075cm}\\
$\mathbf{S}_\mathrm{hist}^{E/F}\leftarrow S_\mathrm{hist}^{r,E/F}\cdot F_{++}\ \ \mathrm{if}\ L_\mathrm{new}^{r,E/F}>\overline{L}_{T3}^{E/F}\land L_\mathrm{new}^{r,E/F}>L_\mathrm{old}^{r,E/F}$\vspace{0.075cm}\\
$\mathbf{S}_\mathrm{hist}^{E/F}\leftarrow -3\ \ \mathrm{if}\ X^r\geq N_\mathbf{X}$\vspace{0.075cm}\\
$\mathbf{S}_\mathrm{hist}^{E/F}\leftarrow -2\ \ \mathrm{if}\ \left[\mathrm{max}\left(0,\ S_\mathrm{hist}^{r,E}-1\right)\right]^2$\\
$\quad\quad\quad\ \ \ +\left[\mathrm{max}\left(0,\ S_\mathrm{hist}^{r,F}-1\right)\right]^2>\left(S_\mathrm{hist}^\mathrm{max}-1\right)^2$\vspace{0.075cm}\\
$\mathbf{S}_\mathrm{hist}^{E/F}\leftarrow -1\ \ \mathrm{if}\ \left\{\mathrm{max}\left[0,\ \left(S_\mathrm{hist}^{r,E}\right)^{-1}-1\right]\right\}^2$\\
$\quad\quad\quad\ \ \ +\left\{\mathrm{max}\left[0,\ \left(S_\mathrm{hist}^{r,F}\right)^{-1}-1\right]\right\}^2>\left[\left(S_\mathrm{hist}^\mathrm{max}\right)^{-1}-1\right]^2$\vspace{0.075cm}\\
$\mathbf{L}_\mathrm{old}^\mathrm{fit,E/F}\leftarrow\mathbf{L}_\mathrm{new}^{E/F}$\vspace{0.075cm}\\
$\overline{L}_\mathrm{new}^{E/F}\leftarrow\mathrm{mean}\left(\mathbf{L}_\mathrm{old}^{\backslash\mathrm{NaN},S_\mathrm{hist}^{r,E}\geq-1,E/F}\right)$\vspace{0.075cm}\\
$\mathrm{if}\ \overline{L}_\mathrm{new}^E\leq \overline{L}_\mathrm{old}^E\land \overline{L}_\mathrm{new}^F\leq \overline{L}_\mathrm{old}^F$\vspace{0.075cm}\\
$\quad\quad p_\mathrm{good}\leftarrow\mathrm{max}\left(0,\ p_\mathrm{good}-\dfrac{p_\mathrm{good}^\mathrm{max}}{N_p}\right)$\vspace{0.075cm}\\
$\mathrm{else}$\vspace{0.075cm}\\
$\quad\quad p_\mathrm{good}\leftarrow\mathrm{min}\left(p_\mathrm{good}+\dfrac{p_\mathrm{good}^\mathrm{max}}{N_p},\ p_\mathrm{good}^\mathrm{max}\right)$\vspace{0.075cm}\\
$\overline{L}_\mathrm{old}^{E/F}\leftarrow\overline{L}_\mathrm{new}^{E/F}$\vspace{0.125cm}\\
\hline
\end{tabular}
\end{table}

To update the selection-determining properties (Algorithm 2), the loss contributions $\mathbf{L}_\mathrm{new}^{E/F}$ of the currently chosen structures are calculated. In addition, loss contributions $\mathbf{L}_{Ti}^{E/F}$ are calculated for which the deviation is scaled by one of four threshold factors $T_i$, with $i={1,2,3,4}$. These vectors are initialized before the first step as $\mathbf{NaN}$. For each of these vectors the mean $\overline{L}_{Ti}^{E/F}$ is determined for the union of available and redundant structures. 

If the new energy or force loss contribution of a structure $r$ is larger than the mean loss contribution applying the largest threshold factor $\overline{L}_{T4}^{E/F}$, then the exclusion strike counter $X^r$ of that structure will be increased by one. The starting value of this counter is zero for each structure. If the loss contributions for energy and forces are lower than the threshold, then the counter will be reset to zero. Afterwards, the adaptive selection factors $S_\mathrm{hist}^{r,E/F}$ are individually updated for energies and forces. If the new loss contribution ${L}_\mathrm{new}^{r,E/F}$ is not lower than the first threshold $\overline{L}_{T1}^{E/F}$, $S_\mathrm{hist}^{r,E/F}$ will have a lower bound of one. An upper bound of one will be applied if the value is not higher than the second threshold $\overline{L}_{T2}^{E/F}$. In this way, only well/badly represented structures can get a small/large $S_\mathrm{hist}^{r,E/F}$ value, while the $S_\mathrm{hist}^{r,E/F}$ values of structures with intermediate representation quality retain around one. Subsequently, $S_\mathrm{hist}^{r,E/F}$ is modified by large and small decrease factors $F_{--}$ and $F_{-}$ as well as small and large increase factors $F_{+}$ and $F_{++}$ depending on the value of the new loss contribution ${L}_\mathrm{new}^{r,E/F}$ compared to the threshold values $\overline{L}_{Ti}^{E/F}$ and the old loss contribution ${L}_\mathrm{old}^{r,E/F}$. These decrease and increase factors are calculated from the hyperparameters $N_{-/--}$ and $N_{+/++}$ by $F_{-/--}=(S_\mathrm{min})^{(N_{-/--})^{-1}}$ and $F_{+/++}=(S_\mathrm{max})^{(N_{+/++})^{-1}}$. Therefore, $N_{-/--}$ and $N_{+/++}$ define the number of repeated applications of the respective factor until $S_\mathrm{min}$ or $S_\mathrm{max}$ is reached. Exceeding these lower and upper bounds leads to the exclusion of the associated structure from training, as described below.

If the exclusion strike counter $X^r$ reaches its limit $N_\mathbf{X}$, $S_\mathrm{hist}^{r,E/F}$ is set to $-3$ and the structure is excluded due to very large errors. If the upper threshold for $S_\mathrm{hist}^{r}$ as defined in Algorithm 2 is exceeded, $-2$ is assigned to $S_\mathrm{hist}^{r,E/F}$ to exclude the structure due to steadily large errors for many training steps. Both assignments mean that the respective training data point is inconsistent with the majority of data. If the lower threshold for $S_\mathrm{hist}^{r}$ is exceeded, $S_\mathrm{hist}^{r,E/F}$ is set to $-1$, classifying a redundant structure, i.e., a structure that has been well represented for many steps. To avoid that too many structures are classified as redundant in the same training step, a maximal fraction of the new redundant training structure per step $p_\mathrm{redun}^\mathrm{max}$ can be set in addition to Algorithm 2. In this way, only 
\begin{align}
N_\mathrm{redun}^\mathrm{max}=\mathrm{max}\left(1,\ \lceil p_\mathrm{redun}^\mathrm{max}\cdot N_\mathrm{fit}\rceil\right)
\end{align}
randomly selected structures of the structures exceeding the lower threshold are classified as redundant and $S_\mathrm{hist}^{r,E/F}$ of the remaining structures is divided by $F_{--}$. This approach reduces the risk that all structures creating some redundancy are by chance excluded in the same step, i.e., none of them remains available for training. 

To update the fraction of good data $p_\mathrm{good}$, the new loss contributions first replace the respective old ones in $\mathbf{L}_\mathrm{old}^{E/F}$. Subsequently, the mean $\overline{L}_\mathrm{new}^{E/F}$ of this vector is calculated for the union of available and redundant training data. If the new means of energies and forces are not larger than the respective old values, $p_\mathrm{good}$ will decrease by $p_\mathrm{good}^\mathrm{max}\,N_p^{-1}$. Otherwise, it increases by the same value. The mean loss contributions for energy and forces are initialized before the first step as infinity. The resulting $p_\mathrm{good}$ has a lower bound of $0$ and an upper bound of $p_\mathrm{good}^\mathrm{max}$. Consequently, $p_\mathrm{good}$ can have $N_p+1$ different values. Finally, $\overline{L}_\mathrm{old}^{E/F}$ is overwritten by $\overline{L}_\mathrm{new}^{E/F}$.

\subsubsection*{Integration of New Data}

For an efficient integration of additional data at a training stage where some structures are already well represented, Algorithm 3 can replace lines 3 to 6 of Algorithm 2. The main idea behind this data integration is that the additional data can have the maximum training probability for several training steps and they cannot be excluded during these steps. In this way, the lMLP can learn the new structures fast, even if the representation quality is very different between the previous and new training data. Without this integration algorithm, the risk of exclusion is high for new data because the typically low errors for the majority of old data lead to a fast increase in the exclusion strike counters and the adaptive selection factors of new data with usually high errors. In addition, the new data initially do not affect the update of the selection-determining properties of the old data to obtain a stable assessment of the representation quality.

\begin{table}[htb!]
\centering
\begin{tabular}{p{8.25cm}}
\hline\vspace{-0.175cm}
\textbf{Algorithm 3:} Integration of new training data.\vspace{0.125cm}\\
\hline\vspace{-0.125cm}
$\mathrm{for}\ i\ \mathrm{in}\ \{1,2,3,4\}$\vspace{0.075cm}\\
$\quad\quad\mathbf{L}_{Ti}^E\leftarrow q\cdot\mathcal{L}\left(\dfrac{T_i\cdot\mathbf{E}^\mathrm{fit}}{\mathbf{N}_\mathrm{atom}^\mathrm{fit}}\right)\ \ \mathrm{if}\ L_\mathrm{old}^{r,E}\neq\mathrm{NaN}$\vspace{0.075cm}\\
$\quad\quad\mathbf{L}_{Ti}^F\leftarrow\dfrac{\mathcal{L}\left(T_i\cdot\mathbf{F^\mathrm{fit}}\right)}{3\mathbf{N}_\mathrm{atom}^\mathrm{fit}}\ \ \mathrm{if}\ L_\mathrm{old}^{r,E}\neq\mathrm{NaN}$\vspace{0.075cm}\\
$\quad\quad\overline{L}_{Ti}^{E/F}\leftarrow\mathrm{mean}\left(\mathbf{L}_{Ti}^{\backslash\mathrm{NaN},S_\mathrm{hist}^{r,E}\geq-1,E/F}\right)$\vspace{0.075cm}\\
$\mathbf{I}\leftarrow I^r+1\ \ \mathrm{if}\ \left(L_\mathrm{new}^{r,E}>\overline{L}_{T2}^E\lor L_\mathrm{new}^{r,F}>\overline{L}_{T2}^F\right)\land L_\mathrm{old}^{r,E}=\mathrm{NaN}$\vspace{0.075cm}\\
$\mathbf{I}\leftarrow 0\ \ \mathrm{if}\ L_\mathrm{new}^{r,E}\leq \overline{L}_{T2}^E\land L_\mathrm{new}^{r,F}\leq \overline{L}_{T2}^F$\vspace{0.075cm}\\
$\mathbf{L}_\mathrm{new}^{E/F}\leftarrow\mathrm{NaN}\ \ \mathrm{if}\ 0<I^r<N_\mathbf{I}$\vspace{0.125cm}\\
\hline
\end{tabular}
\end{table}

In contrast to Algorithm 2, Algorithm 3 does not consider loss contributions of structures evaluated for the first time ($L_\mathrm{old}^{r,E}=\mathrm{NaN}$) in the calculation of the loss contribution threshold values. In this way, we circumvent the effect that most old structures will be considered as relatively well represented just because new structures with typically high errors have been added. Algorithm 3 introduces the integration counter $I^r$, which is initialized as zero for each structure. It is increased by one in every training step of a new structure, which still has an energy or force loss contribution greater than the second lowest threshold. If the latter condition is not satisfied, $I^r$ is set to zero. Subsequently, the loss contribution of new structures with high errors ($I^r>0$) will be reset to $\mathrm{NaN}$ if the integration counter is still below the maximum number of integration steps $N_\mathbf{I}$. This reset retains the classification of a structure to be new and the associated maximum training probability. In this way, new structures are integrated into the general selection process as soon as their accuracy is close to those of the majority of the data, while inconsistent data cannot get stuck in this integration process.

\subsubsection*{Backtracking Loss Gradients of Inconsistent Data}

Many optimizers utilize the momentum of the optimization process to achieve better performance \cite{Eckhoff2024}. For example, Adam \cite{Kingma2015} and the CoRe optimizer \cite{Eckhoff2023, Eckhoff2024} employ the exponential moving average of the loss gradient in the model parameter update to consider the loss gradient history. As a consequence, the loss gradient contributions of inconsistent training data can affect the optimization process even after these data have been excluded from training. To counteract this effect, we propose Algorithm 4 for backtracking loss gradient contributions of inconsistent training data in the CoRe optimizer. This algorithm basically eliminates contributions of training data which have been classified to be inconsistent. We note that Algorithm 4 is optimizer-specific because the optimizer algorithm determines the weight of individual loss gradient contributions in the employed gradient for model parameter updates.

\begin{table}[htb!]
\centering
\begin{tabular}{p{8.25cm}}
\hline\vspace{-0.175cm}
\textbf{Algorithm 4:} Backtracking loss gradients of inconsistent training data $r^\prime$ in the history of the CoRe optimizer.\vspace{0.125cm}\\
\hline\vspace{-0.125cm}
$\mathbf{F}_\mathbf{S}\leftarrow S_\mathrm{hist}^\mathrm{max}\left[\left(S_\mathrm{hist}^\mathrm{max}\right)^{\tfrac{\left(0,1,...,N_{++}-1\right)^\mathrm{T}}{N_{++}}}\right]^{-1}$\vspace{0.075cm}\\
$S_\mathrm{hist}^{\mathrm{max},\mathbf{X}}\leftarrow\left(S_\mathrm{hist}^\mathrm{max}\right)^{\tfrac{N_\mathbf{X}}{N_{++}}}$\vspace{0.075cm}\\
$\mathbf{F}_\mathbf{X}\leftarrow S_\mathrm{hist}^{\mathrm{max},\mathbf{X}}\left[\left(S_\mathrm{hist}^{\mathrm{max},\mathbf{X}}\right)^{\tfrac{\left(0,1,...,N_\mathbf{X}-1\right)^\mathrm{T}}{N_{++}}}\right]^{-1}$\vspace{0.075cm}\\
$\mathbf{P}^\mathrm{sum}\leftarrow\mathbf{S}_\mathrm{hist}^{>0,E}\cdot\mathbf{S}_\mathrm{hist}^{>0,F}\cdot\left[\left(1-p_\mathrm{good}\right)\mathbf{P}_\mathrm{loss}^{S_\mathrm{hist}^{r,E}>0}\right.$\\
$\quad\quad\quad\ \ \left.+p_\mathrm{good}\left(1-\mathbf{P}_\mathrm{loss}^{S_\mathrm{hist}^{r,E}>0}\right)\right]$\vspace{0.075cm}\\
$\mathbf{P}^\mathrm{sum}\leftarrow P_\mathrm{bad}^\mathrm{max}\ \ \mathrm{if}\ P^{\mathrm{sum},r}=\mathrm{NaN}$\vspace{0.075cm}\\
$P^\mathrm{sum}\leftarrow\mathrm{sum}\left(\mathbf{P}^\mathrm{sum}\right)$\vspace{0.075cm}\\
$\mathbf{P}^{r^\prime}\leftarrow\mathbf{F}_{\mathbf{S}\lor\mathbf{X}}\left[\left(1-p_\mathrm{good}\right)P_\mathrm{loss}^{r^\prime}+p_\mathrm{good}\left(1-P_\mathrm{loss}^{r^\prime}\right)\right]$\vspace{0.075cm}\\
$\mathbf{t}_\mathrm{BT}\leftarrow\mathrm{cumsum}\left\{\left[\mathrm{min}\left(\dfrac{N_\mathrm{fit}\cdot \mathbf{P}^{r^\prime}}{P^\mathrm{sum}+\mathbf{P}^{r^\prime}},\ 1\right)\right]^{-1}\right\}$\vspace{0.075cm}\\
$\mathbf{t}_\mathrm{BT}\leftarrow\mathbf{t}_\mathrm{BT}-t_0$\vspace{0.075cm}\\
$\boldsymbol{\upbeta}_1\leftarrow\beta_1^\mathrm{b}+\left(\beta_1^\mathrm{a}-\beta_1^\mathrm{b}\right)\exp\left[-\left(\frac{n_\mathrm{epoch}-\mathbf{t}_\mathrm{BT}}{\beta_1^\mathrm{c}}\right)^2\right]$\vspace{0.075cm}\\
$F_\mathrm{BT}^1\leftarrow\mathrm{sum}\left[\left(1-\boldsymbol{\upbeta}_1\right)\cdot\left(\boldsymbol{\upbeta}_1\right)^{\mathbf{t}_\mathrm{BT}}\right]$\vspace{0.075cm}\\
$F_\mathrm{BT}^2\leftarrow\mathrm{sum}\left[\left(1-\beta_2\right)\cdot\left(\beta_2\right)^{\mathbf{t}_\mathrm{BT}}\right]$\vspace{0.075cm}\\
$L^{r^\prime}\leftarrow \dfrac{q}{N_\mathrm{fit}}\cdot\mathcal{L}\left(\dfrac{E^{r^\prime}}{N_\mathrm{atom}^r}\right)+\dfrac{\mathcal{L\left(\mathbf{F}^{\mathit{r}^\prime}\right)}}{3N_\mathrm{atom}^\mathrm{fit,sum}}$\vspace{0.075cm}\\
$\mathbf{g}\leftarrow\mathbf{g}-F_\mathrm{BT}^1\cdot\dfrac{\partial L^{r^\prime}}{\partial \mathbf{w}}$\vspace{0.075cm}\\
$\mathbf{h}\leftarrow\mathbf{h}-F_\mathrm{BT}^2\cdot\left(\dfrac{\partial L^{r^\prime}}{\partial \mathbf{w}}\right)^2$\vspace{0.075cm}\\
$\mathbf{h}\leftarrow0\ \ \mathrm{if}\ h^\xi<0$\vspace{0.125cm}\\
\hline
\end{tabular}
\end{table}

To determine the weight of the loss gradient contribution for a given structure in the exponential moving average, we need to know in which training steps this structure has been utilized and how large the respective loss function gradient contribution was. To avoid saving this large amount of data for all training data, Algorithm 4 estimates the most probable steps based on a few assumptions and employs the current loss function gradient. First, we assume that a structure is excluded due to errors larger than threshold 3 for $N_{++}$ steps or larger than threshold 4 for $N_\mathbf{X}$ steps. The $S_\mathrm{hist}^{r,E/F}$ value of inconsistent data reveals which of the two cases applies. Moreover, we assume that either only energy errors or only force errors exceed the threshold. To estimate the training probability, all values of the respective adaptive selection factors $\mathbf{F}_\mathbf{S}$ and $\mathbf{F}_\mathbf{X}$ are calculated for both cases. In addition, the non-normalized probabilities $\mathbf{P}^\mathrm{sum}$ need to be determined for all other training data that are still available for training. Therefore, we assume that their current adaptive selection factors and probabilities based on the loss are also a reasonable representation for previous steps, and we estimate the ratio of well and badly represented data by the current $p_\mathrm{good}$ value. $P_\mathrm{bad}^\mathrm{max}$ is utilized if $P^{\mathrm{sum},r}$ equals $\mathrm{NaN}$. In the remaining algorithm, only the sum $P^\mathrm{sum}$ is required.

The following part of Algorithm 4 has to be repeated for each structure $r^\prime$ that has been identified in the current step to be inconsistent. Initially, the non-normalized probabilities $\mathbf{P}^{r^\prime}$ are calculated using the current $P_\mathrm{loss}^{r^\prime}$, $p_\mathrm{good}$, and $\mathbf{F}_\mathbf{S}$ or $\mathbf{F}_\mathbf{X}$. The latter depends on which exclusion reason applies for $r^\prime$. Afterwards, the average training frequency is calculated for each entry in $\mathbf{P}^{r^\prime}$, whereby a structure can only be trained once in a training step. The cumulative sum minus the value of the first vector entry yields the vector of backtracking steps $\mathbf{t}_\mathrm{BT}$. The current epoch $n_\mathrm{epoch}$ minus $\mathbf{t}_\mathrm{BT}$ corresponds to the most probable training steps of structure $r^\prime$ under the aforementioned assumptions. For these steps, the decay hyperparameters $\boldsymbol{\upbeta}_1$ of the CoRe optimizer are calculated. Subsequently, the backtracking factors $F_\mathrm{BT}^1$ and $F_\mathrm{BT}^2$ can be obtained, i.e., the sum of the weights of all loss gradient contributions of structure $r^\prime$ in the steps $n_\mathrm{epoch}-\mathbf{t}_\mathrm{BT}$ can be calculated using the decay hyperparameters $\boldsymbol{\upbeta}_1$ and $\beta_2$.

Finally, we assume that the loss contribution $L^{r^\prime}$ after the model parameter update is also a reasonable representation of the previous loss contributions. We note that recent loss gradients contribute the most, which supports this approximation. We calculate the gradient of $L^{r^\prime}$ with respect to the model parameters $\mathbf{w}$ and its square and subtract these gradients weighted by the respective backtracking factors from the exponential moving averaged loss gradient $\mathbf{g}$ and squared loss gradient $\mathbf{h}$ in the CoRe optimizer. In this way, the loss gradient contributions of structure $r^\prime$ are eliminated. Since this algorithm is approximate, we enforce the required condition that values $h^\xi$ in $\mathbf{h}$ cannot be lower than 0.

Since an optimization step based on energies alone can be much faster than one based on energies and forces (due to the additional differentiation step), a pre-optimization step on energies before the actual optimization step on both energy and forces can increase the training efficiency. To apply Algorithm 4 in this case, $\mathbf{t}_\mathrm{BT}$ needs to be multiplied by two. In addition, the Algorithm has to be repeated for the energy pre-optimization step with the modification that $\mathbf{t}_\mathrm{BT}$ again needs to be multiplied by a factor of two. It also needs to be increased by one, and only the energy loss contribution has to be utilized.

\subsubsection*{Hyperparameters}

For the lADS hyperparameters, we generally recommend the following settings: $p_\mathrm{fit}=\tfrac{1}{30}$, $S_\mathrm{hist}^\mathrm{min}=0.1$, $S_\mathrm{hist}^\mathrm{max}=100$, $T_i=\{0.9, 2.5, 4.0, 7.5\}$, $N_{--}=10$, $N_{-}=30$, $N_{+}=100$, $N_{++}=300$, $N_\mathbf{X}=15$, $p_\mathrm{redun}^\mathrm{max}=0.02$, $p_\mathrm{good}^\mathrm{max}=\tfrac{2}{3}$, $N_p=20$, and $N_\mathbf{I}=30$. The data integration algorithm (Algorithm 3) can be applied as soon as new data are added for the first time. Hence, Algorithm 3 needs to be disabled only if training is started from scratch. In addition, the hyperparameter for balancing the energy and force loss contributions was $q=250$ in this work. These settings were applied unless otherwise stated.

\section{Computational Details}\label{sec:Computational_Details}

\subsection{Exploration of Chemical Reaction Networks}

For the CRN exploration we applied our freely available open-source Software for Chemical Interaction Networks (SCINE) (see Reference \cite{Weymuth2024} for an overview), especially the SCINE modules Chemoton (version 3.0.0) \cite{Unsleber2022, Bensberg2023a}, Molassembler (version 2.0.0) \cite{Sobez2020, Bensberg2023b}, ReaDuct (version 5.0.0) \cite{Vaucher2018, Bensberg2023c}, Puffin (version 1.2.0) \cite{Bensberg2023d}, and Database (version 1.2.0) \cite{Bensberg2023e}. The required DFT calculations were carried out with the quantum chemistry software ORCA (version 5.0.3) \cite{Neese2012, Neese2022}. We executed spin unrestricted DFT calculations with the PBE exchange-correlation functional \cite{Perdew1996} in combination with the def2-TZVP basis set \cite{Weigend2005}. The xTB software \cite{Bannwarth2021} was applied for semi-empirical GFN2-xTB calculations \cite{Bannwarth2019}. For MLP energy, gradient, and Hessian calculations within SCINE, we implemented the module Parrot (version 1.0.0).

To describe the setup of the CRN exploration, we refer here to technical terms in SCINE \cite{Unsleber2020}: A \textit{structure} is a (stationary) point on the potential energy surface and a \textit{compound} is a group of \textit{structures} with the same connectivity as defined in Molassembler. An \textit{elementary step} is a transformation from a minimum energy \textit{structure} to another one through a single transition state. A \textit{reaction} is a group of \textit{elementary steps} connecting the \textit{structures} belonging to two reacting \textit{compounds}. Hence, two \textit{compounds} can be connected by a \textit{reaction}.

\textit{Elementary step} trials in the CRN exploration were set up for one \textit{structure} per \textit{compound}, i.e., we did not explore explicitly changes in the reactivity due to different conformations of a \textit{compound}. However, for bimolecular reaction trials, the required reactive complexes were generated using different attack points of the two \textit{structures} and up to two rotamers of them. Reactive complexes were ignored if the two \textit{structures} were both charged positively or negatively, since electrostatic repulsion makes the resulting \textit{elementary steps} unfavorable. To find the transition state, we employed the Newton Trajectory Algorithm 2 (NT2) (see Reference \cite{Unsleber2022} for details). We allowed uni- and bimolecular reaction trials with one or two intended bond modifications per trial. We note that the resulting number of bond modifications can be lower or higher, since the NT2 scan is a single-ended transition state search algorithm. If an NT2 scan found a transition state, the respective structure was optimized. Afterwards, an intrinsic reaction coordinate (IRC) calculation was carried out to get the reactant and product minimum energy \textit{structures} that belong to the transition state. The spin multiplicity of the resulting \textit{structures} was chosen to be as small as possible. \textit{Structures} that can only be formed by overcoming a single energy barrier larger than $250\,\mathrm{kJ\,mol}^{-1}$ were not considered in the setup of further \textit{elementary step} trials. $10\%$ of the structure conformations that occurred in the NT2 scans and IRC calculations were stored for the benchmark of the lMLPs. In addition, $1\%$ of the structure conformations occurring in the optimizations of reactants, reactive complexes, transition states, and IRC outcomes were saved. The structure conformations were ordered chronologically to enable continual learning retrospectively.

\subsection{Machine Learning Potentials}

We implemented training and prediction of lMLPs in the lMLP software 
(version 2.0.0).
It utilized NumPy (version 1.26.4) \cite{Harris2020}, PyTorch (version 2.3.1) \cite{Paszke2019}, and Numba (version 0.60.0) \cite{Lam2015}.

The eeACSF parameter values of the lMLPs are provided in Tables S1 and S2 in the Supporting Information. Network expressivity by activation rank (NEAR) \cite{Husistein2024} was applied as a training-free pre-estimator of neural network performance to automate the search for the neural network architecture. The resulting architecture contains 135 input neurons, four hidden layers with 117, 137, 164, and 196 neurons, and one output neuron. The training was based on total energies and the atomic force components. The sum of respective reference element energies (Tables S3 and S4 in the Supporting Information) was subtracted from the total energy prior to training. In this work, we trained on PBE/def2-TZVP energies and forces, but, in general, data from any electronic structure method (including dispersion corrections) can be utilized, provided that the potential energy surface is consistent. For each lMLP training, we carried out $20$ independent HDNNP training runs, in which the initial neural network weight parameter values and the reference data assignment to training and test sets were different. The weight parameter initialization was tailored to the activation function $\mathrm{sTanh}(x)\coloneqq1.59223\cdot\tanh(x)$ \cite{Eckhoff2023}. In general, $90\%$ of the reference structures were used for training and $10\%$ for testing.

In the optimization of the lMLP weight parameters, a pre-optimization step was carried out based on the energy-dependent term of the loss function $L_\mathrm{total}$ (Equation \ref{eq:total_loss}), since this term is much faster to evaluate than the force-dependent term. The pre-optimization step employed the energies of all structures that were selected by lADS for the given training epoch. Subsequently, the energies and atomic force components of these structures were utilized in a second optimization step, which included updating the lADS properties. The hyperparameters of lADS matched our recommendations in Section 2.2. To obtain fair comparisons between continual and conventional learning, exceptions were made in Sections 4.3 and 4.4 that are explicitly stated there. The CoRe optimizer (version 1.1.0) \cite{Eckhoff2023, Eckhoff2024, Eckhoff2024a} was applied with the hyperparameters $\beta_1^\mathrm{a}=0.7375$, $\beta_1^\mathrm{b}=0.8125$, $\beta_1^\mathrm{c}=250.0$, $\beta_2=0.99$, $\epsilon=10^{-8}$, $\eta_-=0.55$, $\eta_+=1.2$, $s_\xi^0=10^{-3}$, $s_\mathrm{min}=10^{-6}$, $s_\mathrm{max}=10^{-2}$, $d=0.1$, $t_\mathrm{hist}=250$, and $p_\mathrm{frozen}=0.025$. Exceptions were $d=0.01$ for the weight parameters $\alpha$ and $\beta$ (see Supporting Information S1.1 for the definition of these weight parameters) and $d=0$ for weight parameters associated with the output neuron, since these weight parameters can intentionally have larger values. Additionally, we set $p_\mathrm{frozen}=0$ for these weight parameters.

The prediction was based on an ensemble of the $10$ best individual HDNNPs of all $20$. The respective ranking for this selection was given by the sum of the mean squared errors of the test energies and the test atomic force components, whereby the energy error was scaled by $2\,500\,\text{\AA}^{-2}$ to balance the energy and force values.

\section{Results and Discussion}\label{sec:Results_and_Discussion}

\subsection{Chemical Reaction Network}

A CRN exploration was carried out starting from HCN and H$_2$O. The exploration proceeded in two shells; that is, all initial \textit{elementary step} trials (see Section 3.1 for the definitions of the technical terms of SCINE \cite{Unsleber2020}) were calculated for HCN and H$_2$O, and afterwards, all subsequent trials were calculated for the \textit{compounds} resulting from the initial trials. The resulting CRN is built from $719$ \textit{compounds} and $1\,230$ \textit{reactions}.

The CRN of $\text{HCN}+\text{H}_2\text{O}$ is important in the context of the origin of life \cite{Ferris1984, Ruiz-Mirazo2014}. The initial two steps in the formation of formic acid (up to the formation of formamide) \cite{Das2019a} are contained in the computed CRN. The last step is expected to be absent because only two exploration shells were considered. Other species such as hydrogen isocyanide, methylene imine, aminoacetonitrile, hydrogen peroxide, and molecular hydrogen are included in the CRN, as are many high-energy species. We note that a proper evaluation of this chemistry would require more exploration shells to fully represent the reaction mechanisms.

However, in this work we are mainly interested in a typical ensemble of structures that is encountered in CRN explorations. Therefore, we can limit the exploration to the first two shells. Since the structures that we encounter do not only span minimum-energy and transition-state structures, but also all sorts of positions of the atoms along (not yet optimized) reaction paths, we will call all of them `structure conformations' for the sake of brevity in this work. Some reaction coordinates led to structures of comparatively high energies which have to be represented sufficiently well by the MLP. We compiled fractions of the structure conformations occurring in the \textit{elementary step} trials in a benchmark data set. Details on the number of structure conformations obtained from the different subtasks of the \textit{elementary step} trials are provided in Table S5 in the Supporting Information. In total, $225\,595$ structure conformations were collected from the CRN exploration.

\subsection{Reference Data for the Lifelong Machine Learning Potential}

The CRN benchmark data set can be utilized to validate that an MLP is in principle able to yield the desired chemical accuracy for CRN explorations. Moreover, this large data set allows us to evaluate the performance of continual learning algorithms. These algorithms are necessary for the final goal of rolling explorations of chemical reactivity. The reason is that the exploration may face new reactants, new catalysts, new reductants or oxidants, and so forth, that might not have been well represented by the (initial) training data set. Hence, the lMLP has been designed to efficiently and continuously improve on the fly, with new training data getting absorbed (and redundant old data getting removed) \cite{Eckhoff2023}. As a side remark we note that, without loss of generality for lifelong adaptive data selection, this CRN benchmark dataset only requires the MLP to represent four different chemical elements making it accessible to most MLP descriptors. 

MLPs are typically trained on the difference of the total energy and the sum of reference element energies. The latter can be determined before training on some reference structures. In this work, the reference element energies $E_\mathrm{elem}^\mathrm{ref}$ were calculated from a least squares fit of the total energies of H$_2$, CH$_4$, NH$_3$, and H$_2$O as a function of their stoichiometries. These reference element energies are applied in Table \ref{tab:reference_data} to show the range and standard deviation of the energies without atomic contributions for the benchmark data.

\begin{table}[htb!]
\caption{Numbers of structure conformations $N_\mathrm{conf}$, as well as ranges and standard deviations of energies $E-E_\mathrm{elem}^\mathrm{ref}$ and atomic force components $F_{\alpha,n}$ for the different data sets.}
\begin{center}
\begin{tabular}{lrr}
\hline\hline\vspace{-0.325cm}\\
& \multicolumn{1}{l}{$\text{PBE}-\text{GFN2}$} & \multicolumn{1}{l}{PBE}\vspace{0.075cm}\\
\hline\hline\vspace{-0.325cm}\\
$N_\mathrm{conf}$ & $225\,543$ & $225\,045$\vspace{0.075cm}\\
\hline\vspace{-0.325cm}\\
$\left(E-E_\mathrm{elem}^\mathrm{ref}\right)^\mathrm{range}\,/\,\mathrm{meV\,atom}^{-1}$ & $1\,578.2$ & $4\,076.5$\\
$\left(E-E_\mathrm{elem}^\mathrm{ref}\right)^\mathrm{std}\,/\,\mathrm{meV\,atom}^{-1}$ & $61.0$ & $131.0$\vspace{0.075cm}\\
\hline\vspace{-0.325cm}\\
$F_{\alpha,n}^\mathrm{range}\,/\,\mathrm{meV}\,\text{\AA}^{-1}$ & $8\,717$ & $29\,913$\\
$F_{\alpha,n}^\mathrm{std}\,/\,\mathrm{meV}\,\text{\AA}^{-1}$ & $649$ & $448$\vspace{0.075cm}\\
\hline\hline
\end{tabular}
\end{center}
\label{tab:reference_data}
\end{table}

To further increase the transferability and accuracy of the lMLP, we train the lMLP on the energy difference of PBE DFT energies and GFN2 semi-empirical energies. In this approach, which is inspired by AIQM1 \cite{Zheng2021}, the computational efficiency is still at an affordable level. We note that the semi-empirical method can be replaced by another fast base model. However, the method should not restrict the application range (by contrast to what hampers, for instance, non-reactive force fields), and the difference in the potential energy surface should not be too large, since otherwise the benefit vanishes. For example, a very simple Mie potential with a cutoff radius and parameters trained on DFT data did not improve the results. The advantage of this $\Delta$-learning approach can be seen in Table \ref{tab:reference_data} already because the energy and force ranges are significantly smaller than those of the pure PBE data. Smaller ranges can facilitate the achievement of higher absolute prediction accuracy, since the same relative error leads to a smaller absolute error.

For the PBE data, structures with absolute atomic force components larger than $15\,\mathrm{meV}\,\text{\AA}^{-1}$ were disregarded in lMLP training and performance evaluation, as the accurate prediction of highly unstable structures is not relevant in the application for CRNs. We note that this criterion does not exclude high energy structures per se. For example, the forces of transition states are per definition zero. Since we consider force differences in the $\text{PBE}-\text{GFN2}$ data, we had to adjust the value of the criterion for these data. As our intension is to show that $\Delta$-learning can improve the accuracy, we chose a value that leads to a wider range in terms of structure stability. In this way, representing the data with the same accuracy is more difficult. Therefore, we adjusted the threshold to $4.45\,\mathrm{meV}\,\text{\AA}^{-1}$ for $\text{PBE}-\text{GFN2}$ data, excluding less structures for $\text{PBE}-\text{GFN2}$ data than for PBE data.

\subsection{Continual Learning vs.\ Iterative Learning}

This and the following two sections evaluate the performance of lMLPs in different aspects: In this section, we compare the results of the continual learning algorithms to iterative learning. In this way, we can demonstrate the advantage of lMLPs to conventional MLPs in rolling CRN explorations. In the next section, we analyze the accuracy and efficiency gain obtained by lADS compared to random data selection, which are caused by training data focus and reduction. In Section 4.5, we validate that the lMLP can reach the target accuracy for CRN energies.

In an lMLP-driven CRN exploration, the lMLP is pre-trained on an initial data set and then continuously trained on additional, initially unknown data that are flowing in. We constructed a reproducible setting for this exploration process in the following way: We ordered the CRN benchmark data by their occurrence during the DFT exploration process and split them into $N_\mathrm{data}$ equally sized sets. Hence, the structures in the initial set represent reactions between HCN and H$_2$O, while those of the final sets contain reactions between larger and more complex molecules. For simplicity, we started the performance evaluation by training an lMLP from scratch on the first set and then continuously trained one set after the other, utilizing all training data from each set. This training process was monitored and analyzed. We note that the continual learning could have been started from a pre-trained lMLP as well.

\begin{figure}[tb!]
\centering
\includegraphics[width=\columnwidth]{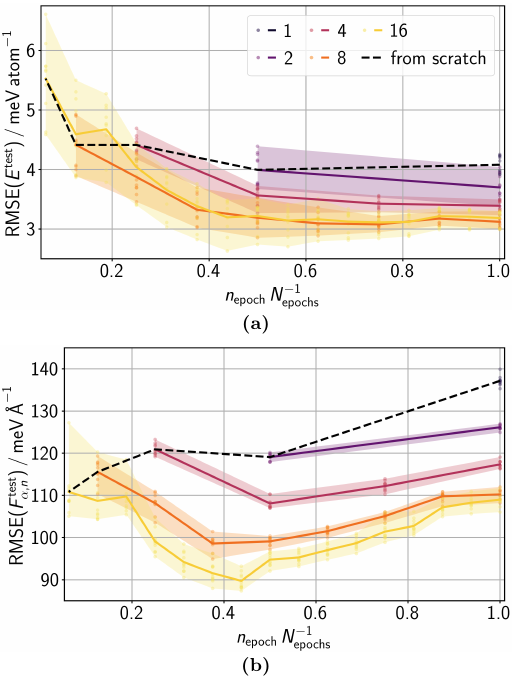}
\caption{Test RMSEs of \textbf{(a)} energies $E^\mathrm{test}$ and \textbf{(b)} atomic force components $F_{\alpha,n}^\mathrm{test}$ after training $1\,000$ epochs on the (extended) data. Trainings were carried out on a sequence of $1$, $2$, $4$, $8$, and $16$ data sets. The total number of epochs $N_\mathrm{epochs}$ is higher for training in more sets, resembling the exploitation of previously acquired knowledge. A constant fraction of $\tfrac{1}{30}$ of all training structures in the respective data set was utilized per epoch. In this way, we can compare results obtained with the same number of structure evaluations, whereby we do not count the evaluations required for training the previous MLP. To avoid instabilities in lADS due to the missing adaption of the number of fitted structures per epoch, we applied here $T_1=0.75$, $N_{-}=40$, $N_{++}=400$, and $p_\mathrm{redun}^\mathrm{max}=0.015$. In this figure and in Figures \ref{fig:lMLP_constant} and \ref{fig:lMLP_random}, $n_\mathrm{epoch}\,N_\mathrm{epochs}^{-1}$ represents a relative scale for the learning curves on the test data. In this figure, the number of underlying training data coincides for the graphs when dots are plotted at the given value of $n_\mathrm{epoch}\,N_\mathrm{epochs}^{-1}$ for these graphs. These dots represent RMSEs of individual HDNNP ensemble members, lines show their mean, and shaded areas span their range. The black dashed line represents the mean RMSE of training from scratch.}\label{fig:lMLP_full_final}
\end{figure}

First, we show how much the continuation of lMLP training can improve the accuracy/cost ratio compared to starting the training after each data addition from scratch again. Therefore, we split the CRN data into $N_\mathrm{data}=\{1,2,4,8,16\}$ sets and trained each set for $1\,000$ epochs. Hence, for a larger number of sets, the lMLP is trained in total for more epochs. This scenario resembles the frequently occurring case in which an MLP needs to be improved during a study and the previous MLP can be employed at no additional cost. Therefore, training for the same number of epochs with the same constant fraction of structures provides a comparison of learning from scratch and continual learning at equal effective cost. The accuracy of the test data after the initial training from scratch for each number of sets is highlighted in Figures \ref{fig:lMLP_full_final} (a) and (b). The accuracy of continued training is consistently and significantly better than that of training from scratch for the same underlying training data. This trend highlights the benefit of continual learning compared to conventional iterative learning because it confirms that the previously learned expertise of the model can be exploited. The accuracy appears to converge to a lower limit for training in more sets.

\begin{figure}[htb!]
\centering
\includegraphics[width=\columnwidth]{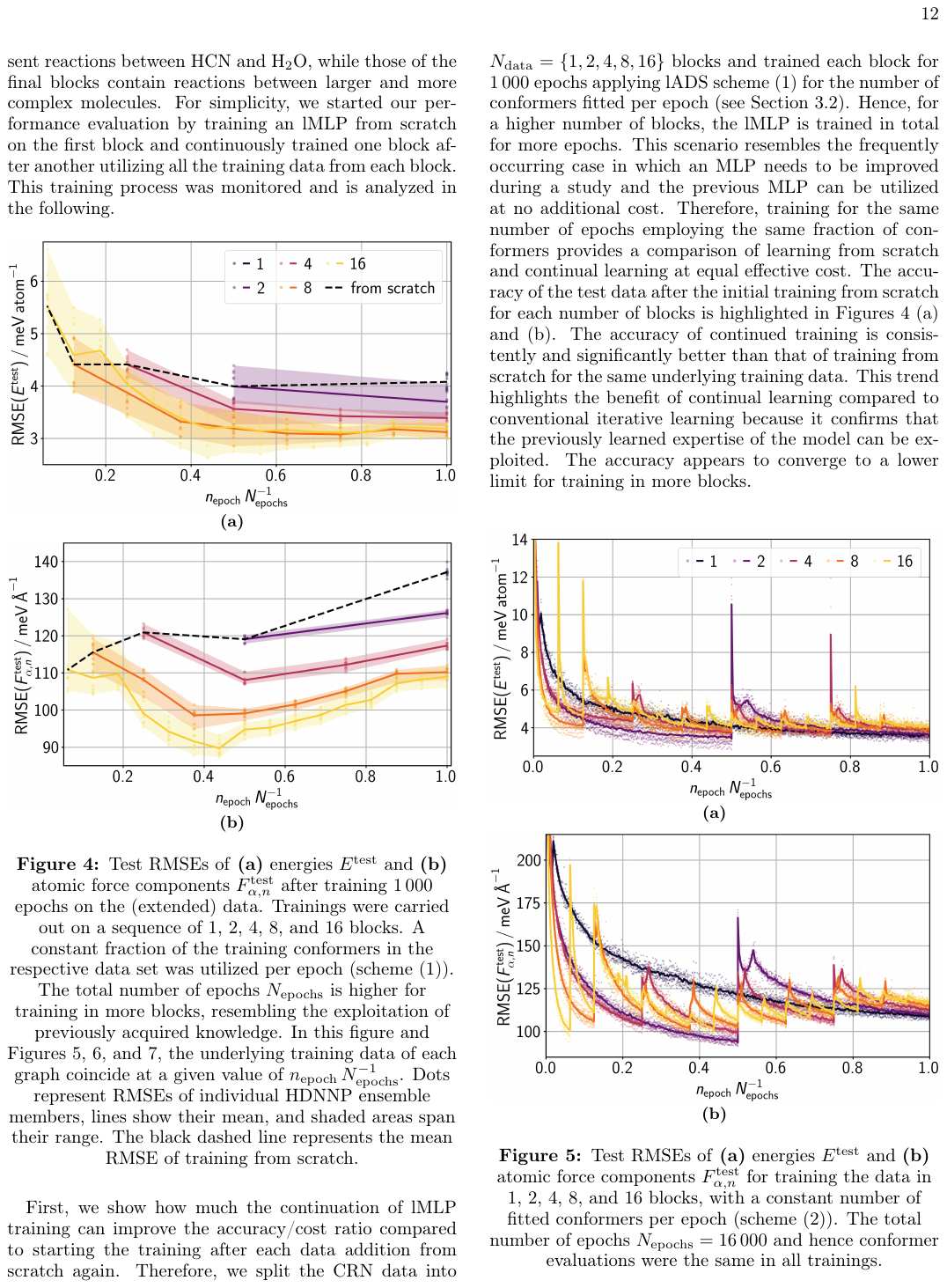}
\caption{Test RMSEs of \textbf{(a)} energies $E^\mathrm{test}$ and \textbf{(b)} atomic force components $F_{\alpha,n}^\mathrm{test}$ for training the data in $1$, $2$, $4$, $8$, and $16$ sets. All these trainings performed the same number of structure evaluations to obtain a fair comparison. Out of this reason, we replaced the adaption of the number of fitted structures per epoch in lADS by a constant number of $N_\mathrm{fit}=750$ structures and trained in total for $N_\mathrm{epochs}=16\,000$ epochs. We readjusted $p_\mathrm{redun}^\mathrm{max}$ to $0.0125$ due to the change in $N_\mathrm{fit}$. Spikes in the graphs originate from data additions. The dots represent RMSEs of individual HDNNP ensemble members and lines show their mean.}\label{fig:lMLP_constant}
\end{figure}

The simplest learning case is training on all structures from the start, i.e., a stationary batch of data. However, this case is unfeasible in many applications that often require active learning and/or subsequent tasks emerge during the application's progress which have additional training data demands. Hence, additional data sets or even a continuous stream of new data need to be learned after the initial training phase. However, training on all data from the start provides a reference for the maximally achievable accuracy of the lMLP, so that we can assess the quality of the results obtained with continual learning. Hence, we trained lMLPs on different numbers of data sets ($N_\mathrm{data}=\{1,2,4,8,16\}$) to go from this simplest learning case to more and more continual learning. To compare at the same absolute cost, the total number of structure evaluations in the training was the same for all cases. Figures \ref{fig:lMLP_constant} (a) and (b) show that the final test errors increase only slightly with more sets of data. Data addition is visible by spikes in the test RMSE, as the new test data can deviate significantly from previously trained data and can therefore lead to large errors. Recovery of accuracy in a small number of steps demonstrates good integration of the additional data. The peak height reduces with more data since the fraction of new data decreases and the probability increases that the necessary information has already been (partially) trained. In conclusion, incremental or continual learning does not yield the same accuracy as training on all data from the beginning as expected. However, in many practical applications, the large cost reduction of continual learning is more important than the small error increase compared to iterative learning that starts for each data addition from scratch again.

\subsection{Lifelong Adaptive Data Selection}

\begin{figure}[htb!]
\centering
\includegraphics[width=\columnwidth]{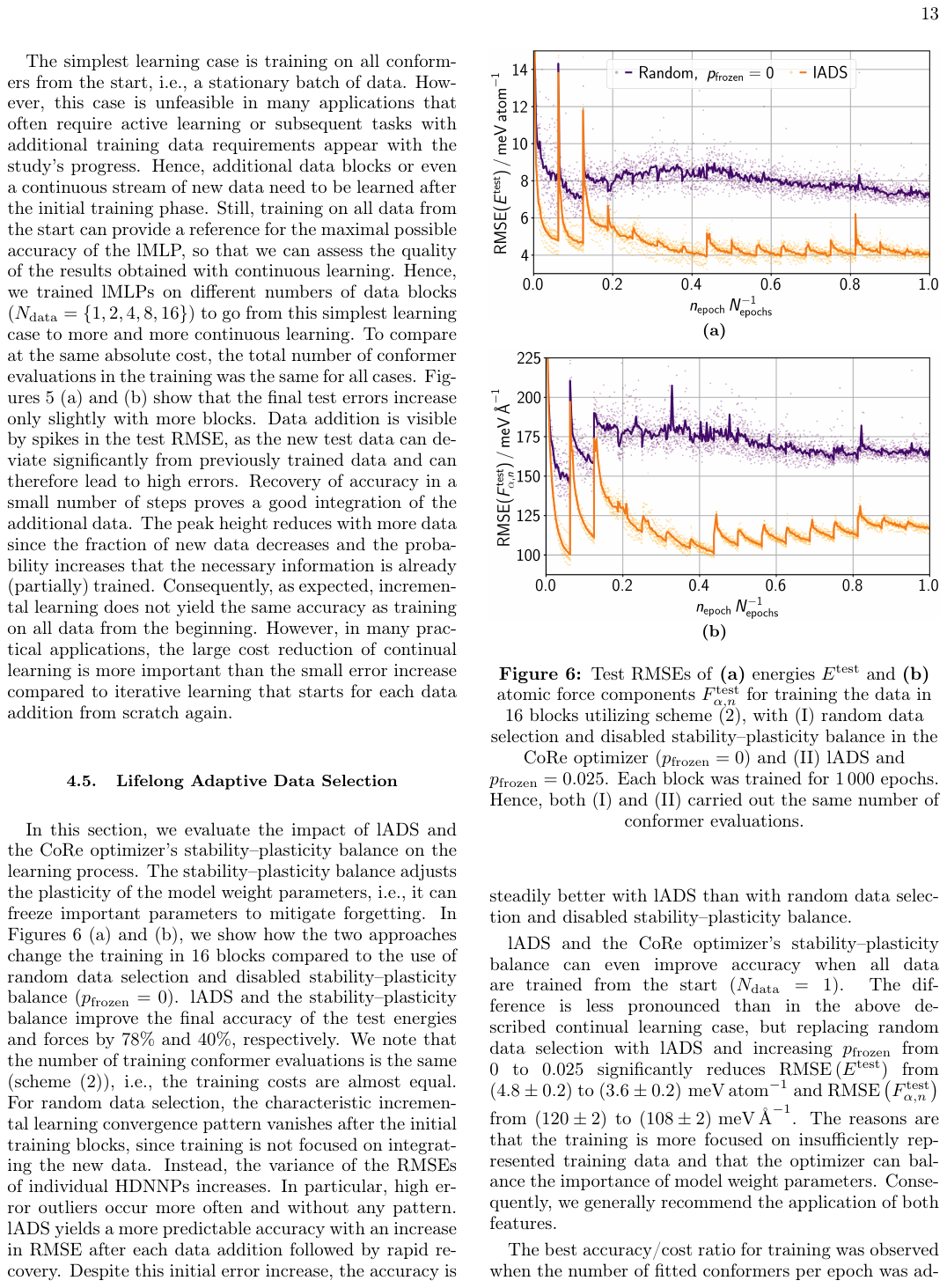}
\caption{Test RMSEs of \textbf{(a)} energies $E^\mathrm{test}$ and \textbf{(b)} atomic force components $F_{\alpha,n}^\mathrm{test}$ for training the data in $16$ sets, with (I) random data selection and disabled stability--plasticity balance in the CoRe optimizer ($p_\mathrm{frozen}=0$) and (II) lADS and $p_\mathrm{frozen}=0.025$. Analogous to Figure \ref{fig:lMLP_constant}, a constant number of fitted structures per epoch (including the readjustment of $p_\mathrm{redun}^\mathrm{max}$) was employed for a fair comparison based on the same number of structure evaluations. Each of the $16$ sets was trained for $1\,000$ epochs. The dots represent RMSEs of individual HDNNP ensemble members and lines show their mean.}\label{fig:lMLP_random}
\end{figure}

In this section, we study (I) the effect of continual learning approaches (lADS and the stability--plasticity balance of the CoRe optimizer) on the learning process and (II) the reduction of training data obtained by lADS. The stability--plasticity balance of the CoRe optimizer can freeze important model weight parameters to mitigate forgetting. In Figures \ref{fig:lMLP_random} (a) and (b), we show how lADS and the stability--plasticity balance change the learning curve on test data for training in $16$ sets compared to the use of random data selection and disabled stability--plasticity balance ($p_\mathrm{frozen}=0$). The two continual learning approaches improve the final accuracy of the test energies and forces by $78\%$ and $40\%$, respectively. We note that the number of training structure evaluations is the same so that the training costs are almost identical. For random data selection, the characteristic incremental learning convergence pattern vanishes after the initial training sets, since training is not focused on integrating the new data. Instead, the variance of the RMSEs of individual HDNNPs increases. In particular, large-error outliers occur more often, following no pattern. lADS yields a more predictable accuracy with an increase in RMSE after each data addition followed by rapid recovery. Despite this initial error increase, the accuracy is steadily better with lADS than with random data selection and disabled stability--plasticity balance.

lADS and the stability--plasticity balance can even improve the accuracy if all data are trained from the start ($N_\mathrm{data}=1$). The difference is less pronounced than in the continual learning case, but replacing random data selection with lADS and increasing $p_\mathrm{frozen}$ from $0$ to $0.025$ significantly reduces $\mathrm{RMSE}\left(E^\mathrm{test}\right)$ from $\left(4.8\pm0.2\right)$ to $\left(3.6\pm0.2\right)\,\mathrm{meV\,atom}^{-1}$ and $\mathrm{RMSE}\left(F_{\alpha,n}^\mathrm{test}\right)$ from $\left(120\pm2\right)$ to $\left(108\pm2\right)\,\mathrm{meV}\,\text{\AA}^{-1}$. The reasons for this effect are that the training is more focused on insufficiently represented training data and that the optimizer can balance the importance of model weight parameters. Consequently, we generally recommend the application of both features.

\begin{figure}[tb!]
\centering
\includegraphics[width=\columnwidth]{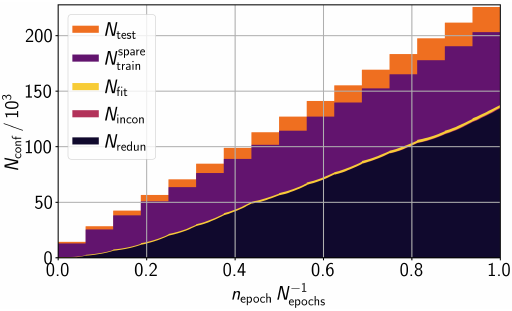}
\caption{Total number of reference structure conformations $N_\mathrm{conf}$ as a function of the training progress $n_\mathrm{epoch}N_\mathrm{epochs}^{-1}$. The colors represent the fractions of test data $N_\mathrm{test}$, not employed $N_\mathrm{train}^\mathrm{spare}$ and employed training data $N_\mathrm{fit}$, and training data disregarded due to inconsistencies $N_\mathrm{incon}$ and due to redundancy $N_\mathrm{redun}$. Data assignment to the classes of being employed or not being employed in fitting can interchange in each epoch, while assignments to test or disregarded data are final. The number of fitted structures per epoch was adjusted to the number of training structures that were not sorted out (as in Algorithm 1).}\label{fig:lMLP_data_assignment}
\end{figure}

The best accuracy/cost ratio for training was observed when the number of fitted structures per epoch was adjusted based on a constant fraction of the training structures that were still employed for training. This adjustment is also important for a stable and well-balanced assignment of structures to be redundant in training. Figure \ref{fig:lMLP_data_assignment} shows the evolution of the data assignments for training in $16$ sets, whereby each set was trained for $1\,000$ epochs ($(26.7\pm0.2)\cdot10^6$ force calculations in total). A steady increase of the number of structures sorted out can be observed. In each of the $10$ individual HDNNP trainings, $134\,452\pm511$ or $(66.2\pm0.3)\%$ of the $202\,986$ training structures were in the end classified as redundant. Only $810\pm43$ or $(0.40\pm0.02)\%$ of the training structures were classified as inconsistent. Hence, the DFT data appear to be of good quality, and the lMLP is able to represent the majority of these data. Consequently, $67\,725\pm538$ or $(33.4\pm0.3)\%$ of the structures remain for rehearsal in further continued training. In this way, a significant speed-up can be provided compared to training again on all data for mitigating catastrophic forgetting. The reliability of the lADS approach is confirmed by the small variances in these assignments for the different individually trained HDNNPs of the ensemble.

Figures S1 (a) and (b) in the Supporting Information show that the accuracy for the test data after training each data addition remains at a similar level, despite that about two thirds of the training data have been removed. This trend confirms that training data classified as redundant is indeed not needed to retain previous expertise, since the latter is required to obtain low test errors. Moreover, due to the removal of training structures, fewer structures are fitted per epoch. With a decreasing number of added data, the training focuses on smaller data subsets, leading to fast data classifications. In this way, training becomes more efficient and continual learning of small amounts of additional data becomes reasonable. For training only on few additional data, the number of epochs per data addition can be reduced, since less information needs to be integrated. Due to Algorithm 3, these few new data points are still in the focus of training enabling efficient integration. 

\subsection{Lifelong Machine Learning Potentials for Exploration of Chemical Reaction Networks}

\begin{figure}[htb!]
\centering
\includegraphics[width=\columnwidth]{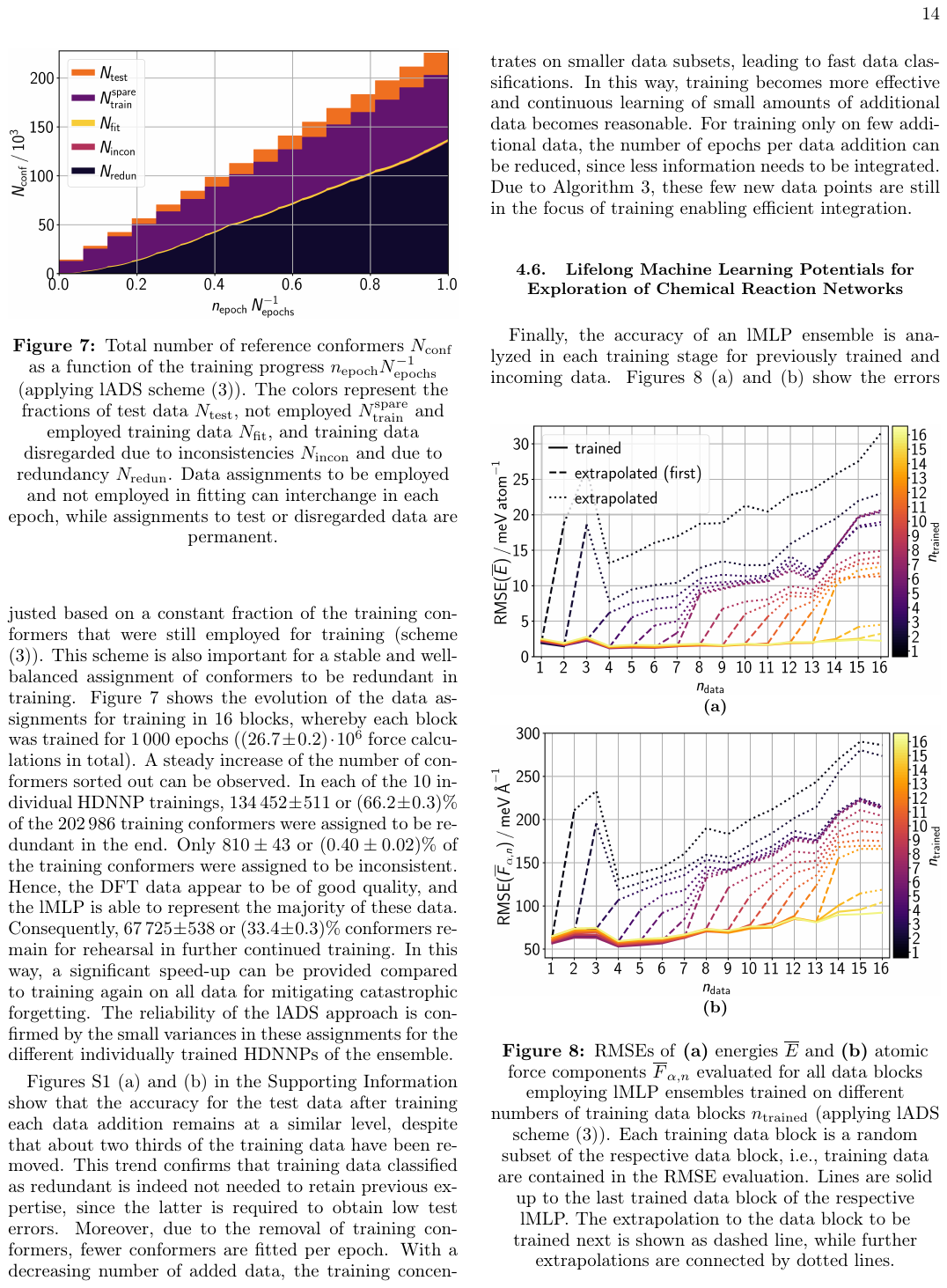}
\caption{RMSEs of \textbf{(a)} energies $\overline{E}$ and \textbf{(b)} atomic force components $\overline{F}_{\alpha,n}$ evaluated with lMLP ensembles at different training stages. The total CRN benchmark data was split into $16$ chronologically ordered sets of equal size, with indices $n_\mathrm{data}=\{1,...,16\}$. The lMLP ensembles were trained on the sets $n_\mathrm{data}=1$ to $n_\mathrm{trained}$. The training data set of each lMLP ensemble member was a random subset of the respective sets. The RMSE evaluation of the lMLP ensembles was based on all data of the respective set $n_\mathrm{data}$ and hence contained trained data. Lines are solid up to the last trained data set of the respective lMLP ensemble. The extrapolation to the data set to be trained next is shown as dashed line, while further extrapolations are connected by dotted lines. In this figure and in Figure \ref{fig:lMLP_accuracy}, lines are shown to guide the eye, but only values at integer numbers of $n_\mathrm{data}$ are meaningful.}\label{fig:lMLP_performance_monitoring}
\end{figure}

Finally, the accuracy of an lMLP ensemble is analyzed in each training stage for previously trained and incoming data. Figures \ref{fig:lMLP_performance_monitoring} (a) and (b) show the errors for each number of trained data sets $n_\mathrm{trained}$ in our simulation of a rolling CRN exploration with data sets $n_\mathrm{data}=\{1,...,16\}$. In general, the lMLP shows in every training stage very good accuracy for the data on which it was already trained. The energy accuracy is relatively constant for all trained data sets. The force RMSE increases slightly with increasing $n_\mathrm{data}$. However, some variation is expected because, on the one hand, more data needs to be well represented and the complexity of the structures increases with $n_\mathrm{data}$, while, on the other hand, the model architecture stays constant. Hence, the model could initially be too complex for the data and finally be affected by capacity issues.

We point out that the lMLP accuracy for the initial data remains almost constant with an increasing number of training stages. Only for the forces, a slight increase in error is notable. However, as the error also slightly increases for additionally trained data, the reason can be a capacity issue due to the constant model architecture. Consequently, the small subset of training data chosen by lADS is sufficient to retain previous expertise. Predictions of data to be trained in the upcoming training stages result in higher errors, highlighting the efficacy of training. The more data sets are in between the last trained set and the evaluated untrained set, the higher the error. The reason is the chronological ordering of the CRN data that leads to an increasing difference in the data, since the chemical reaction process progresses. Hence, the lMLP can efficiently and continuously learn additional data, while previous expertise is kept. Therefore, the lMLP approach is applicable for a rolling exploration. To produce efficient and accurate models, even when training is continued for a large number of additional data points, we expect algorithms to come into play that adjust (grow and shrink) the model architecture during training \cite{Yoon2018}.

To point out the benefit of the continual learning algorithms, we can compare Figures \ref{fig:lMLP_performance_monitoring} (a) and (b) to Figures S2 (a) and (b) in the Supporting Information. The latter figures show the results for the same training task, but without applying rehearsal of training data, lifelong adaptive data selection, and the stability plasticity balance of the CoRe optimizer. Hence, we only trained on the added data starting from the previously obtained model weight parameters. In this case, the MLP ensembles also show very good accuracy for the respective last trained data set. However, the error increases a lot, the more data sets are in between the last trained set and the evaluated set. This trend does not only apply to upcoming training data sets, but also to data sets that were already trained. For example, the MLP ensemble yields the highest RMSE value for the first trained data set after training on all data sets compared to any previous training stage. The value of this RMSE is even similar to the RMSE value on the last data set of the MLP ensemble that was only trained on the first data set. Therefore, catastrophic forgetting occurs in this training task when we do not apply continual learning. If all previous training data are used in every training, we will also avoid catastrophic forgetting. However, this approach will lead to much higher computational demand compared to continual learning. We note that a sequence of several fine-tunings of a foundation model in a row can lead to similar forgetting if continual learning is not applied. Hence, even the expertise on previously fine-tuned data may be forgotten after a sequence of model fine-tunings.

\begin{figure}[htb!]
\centering
\includegraphics[width=\columnwidth]{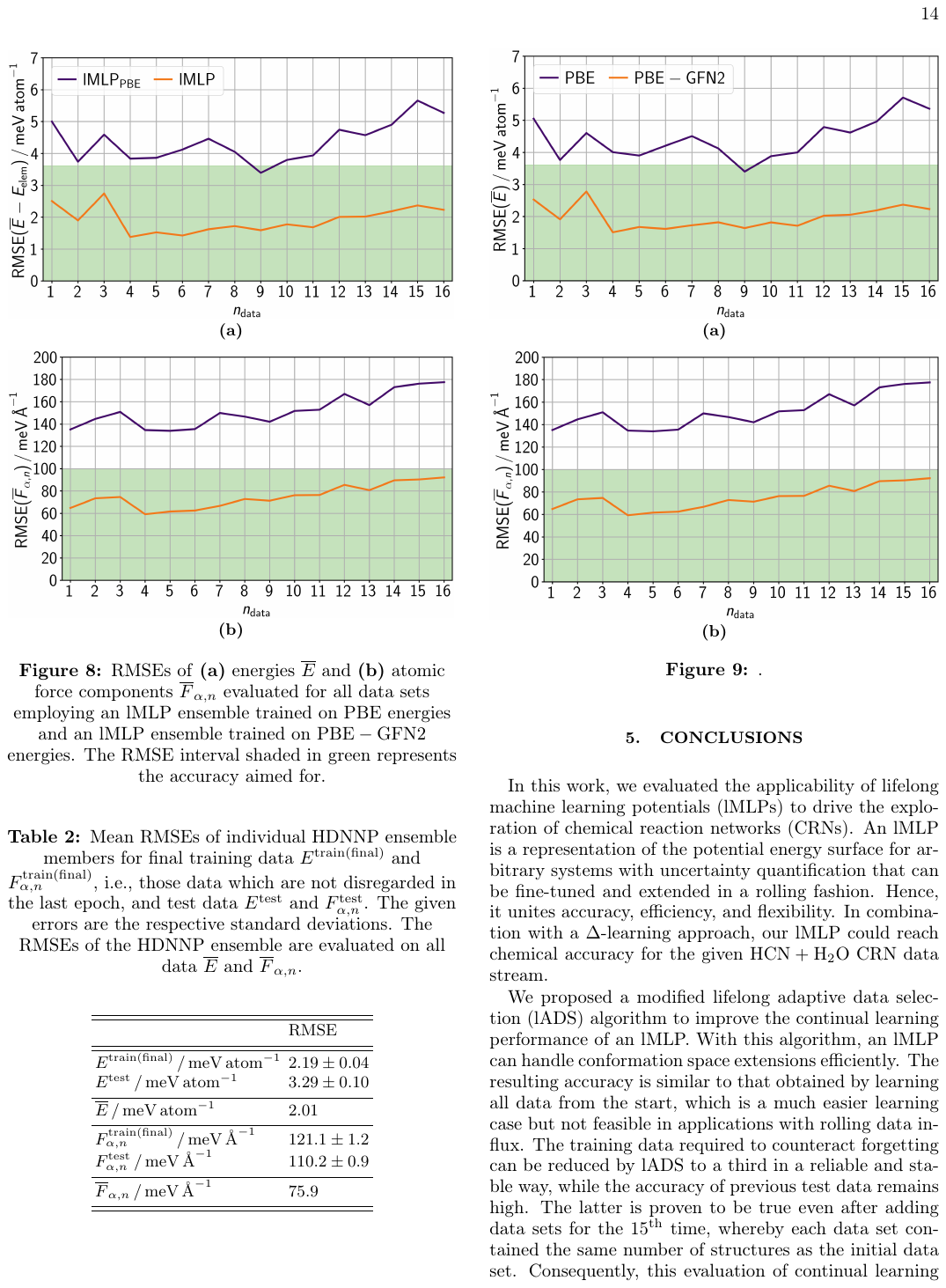}
\caption{RMSEs of \textbf{(a)} energies $\overline{E}$ and \textbf{(b)} atomic force components $\overline{F}_{\alpha,n}$ evaluated for all data sets $n_\mathrm{data}=\{1,...,16\}$ employing an lMLP ensemble trained on PBE data and an lMLP ensemble trained on $\text{PBE}-\text{GFN2}$ data. The RMSE interval shaded in green represents the accuracy aimed for.}\label{fig:lMLP_accuracy}
\end{figure}

To enable reliable kinetic modeling based on the CRN energies, we require at least chemical accuracy, i.e., $1\,\mathrm{kcal\,mol}^{-1}=4.184\,\mathrm{kJ\,mol}^{-1}$, or better. Since the benchmark data contains systems with up to 12 atoms, a minimum energy accuracy of about $3.614\,\mathrm{meV\,atom}^{-1}$ must be the target. Figure \ref{fig:lMLP_accuracy} (a) reveals that the lMLP ensemble accuracy is below this target threshold. Similarly, the MLP accuracy target of $100\,\mathrm{meV}\,\text{\AA}^{-1}$ for the atomic force components \cite{Behler2021} is satisfied (Figure \ref{fig:lMLP_accuracy} (b)). In addition, Figures \ref{fig:lMLP_accuracy} (a) and (b) show the advantage of the $\Delta$-learning approach. If an lMLP is trained in the same way on pure PBE data, the resulting energy and force RMSEs will be approximately twice as large as those of the lMLP trained on $\text{PBE}-\text{GFN2}$ data. This reduction in error is required to yield chemical accuracy for this CRN data set with the given MLP base method.

\begin{table}[htb!]
\caption{RMSEs of individual HDNNP ensemble members and the HDNNP ensemble (trained on $\text{PBE}-\text{GFN2}$ data). For the individual HDNNP ensemble members, the mean and standard deviation of every RMSE are given for the final training data $E^\mathrm{train(final)}$ and $F_{\alpha,n}^\mathrm{train(final)}$, i.e., those data which were not disregarded in the last epoch, and test data $E^\mathrm{test}$ and $F_{\alpha,n}^\mathrm{test}$. For the HDNNP ensemble, the RMSEs were evaluated on all data $\overline{E}$ and $\overline{F}_{\alpha,n}$.}
\begin{center}
\begin{tabular}{ll}
\hline\hline\vspace{-0.325cm}\\
& $\mathrm{RMSE}$\vspace{0.075cm}\\
\hline\hline\vspace{-0.325cm}\\
$E^\mathrm{train(final)}\,/\,\mathrm{meV\,atom}^{-1}$ & $2.19\pm0.04$\\
$E^\mathrm{test}\,/\,\mathrm{meV\,atom}^{-1}$ & $3.29\pm0.10$\vspace{0.075cm}\\
\hline\vspace{-0.325cm}\\
$\overline{E}\,/\,\mathrm{meV\,atom}^{-1}$ & $2.01$\vspace{0.075cm}\\
\hline\vspace{-0.325cm}\\
$F_{\alpha,n}^\mathrm{train(final)}\,/\,\mathrm{meV}\,\text{\AA}^{-1}$ & $121.1\pm1.2$\\
$F_{\alpha,n}^\mathrm{test}\,/\,\mathrm{meV}\,\text{\AA}^{-1}$ & $110.2\pm0.9$\vspace{0.075cm}\\
\hline\vspace{-0.325cm}\\
$\overline{F}_{\alpha,n}\,/\,\mathrm{meV}\,\text{\AA}^{-1}$ & $75.9$\vspace{0.075cm}\\
\hline\hline
\end{tabular}
\end{center}
\label{tab:lMLP_accuracy}
\end{table}

The results in Figures \ref{fig:lMLP_performance_monitoring} and \ref{fig:lMLP_accuracy} include the evaluation of training data (in contrast to Figures \ref{fig:lMLP_full_final} to \ref{fig:lMLP_random}). For an unbiased evaluation, pure test data need to be employed which are available for the individual HDNNPs of the lMLP ensemble. Table \ref{tab:lMLP_accuracy} shows that the mean RMSE of the test energies also satisfies the chemical accuracy criterion. The mean RMSE of the test atomic force components is close to the target accuracy. Since the ensemble results are, in general, better than the individual ensemble member predictions, the force RMSE is sufficiently small. We note that the RMSE of the final training data set can be larger than the test RMSE because about two thirds of well represented data were sorted out from the training data set during training, while the test data set was randomly chosen before training and remained unchanged.

\section{Conclusions}

In this work, we evaluated the applicability of lifelong machine learning potentials (lMLPs) to drive the exploration of chemical reaction networks (CRNs). An lMLP is a representation of the potential energy surface for arbitrary systems with uncertainty quantification that can be fine-tuned and extended in a rolling fashion. Hence, it unites accuracy, efficiency, and flexibility. In combination with a $\Delta$-learning approach, our lMLP can reach chemical accuracy for the given $\text{HCN}+\text{H}_2\text{O}$ CRN data stream.

We proposed a modified lifelong adaptive data selection (lADS) algorithm to improve the continual learning performance of an lMLP. With this algorithm, an lMLP can handle conformation space extensions efficiently. The resulting accuracy is similar to that obtained by learning all data from the start, which is a much easier learning case but not feasible in applications with rolling data influx. The training data required to counteract forgetting can be reduced by lADS to a third in a reliable and stable way, while the accuracy of previous test data remains high. The latter is proven to be true even after adding data sets for the 15$^\text{th}$ time, whereby each data set contained the same number of structures as the initial data set. Consequently, this evaluation of continual learning performance is significantly beyond our initial proof for one single addition of data \cite{Eckhoff2023}. We note that due to adjustable training probabilities for each data point, the integration of added data is more efficient than in plain training of added data and a third of the previously trained data. Moreover, our results confirm that continual learning is able to take advantage of already acquired expertise to improve the final accuracy compared to training from scratch for the same number of epochs. Furthermore, we found that lADS and the continual learning features of the CoRe optimizer can improve the final accuracy not only in continual learning but also in learning stationary data.

Consequently, this work can be considered a proof of principle that lMLPs have all attributes to explore CRNs in a rolling fashion. Hence, lMLPs can be reliably applied on-the-fly during an exploration, where the CRN will be generated directly with an lMLP (instead of DFT as in this work). Based on the uncertainty quantification (accessible in an MLP ensemble approach) it is then possible to decide on where additional DFT data need to be generated for the refinement of the lMLP in subsequent lifelong learning epochs. Uncertain results can then be either directly replaced by the DFT data or recalculated by the improved lMLP afterwards. 

To start such a process from a reasonable initial lMLP, the lMLP can be pre-trained on a large and diverse data set such as one of those employed for foundation models or universal MLPs. We note that the lMLP concept can be applied to other MLP methods than HDNNPs as well, enabling the usage of an already pre-trained foundation model as initial lMLP parametrization. With the increasing size of the data sets used for pre-training of foundation models \cite{Levine2025} as well as the increasing number of diverse lMLP applications, the demand for such initial training events becomes rarer. By community efforts, we can head towards lMLPs that are generally applicable out-of-the-box. However, due to the enormous size of the chemical space, this is a long way to go. Therefore, continual learning is required to train efficiently on unknown or insufficiently represented structures occurring in the actual simulations of interest. In this way, the approach is similar to transfer learning or fine-tuning of foundation models on system-specific data, but continual learning harbors the advantage that learning can be continued for much more than one iteration.

\section*{Data Availability}

The data underlying this study and related self-written software are openly available on Zenodo \cite{Eckhoff2025, Eckhoff2025a}.

\section*{Code Availability}

The lMLP software will be available on GitHub (\url{https://github.com/ReiherGroup/lMLP}) and PyPI (\url{https://pypi.org/project/lmlp}.

The SCINE Parrot software module will be available
on GitHub (\url{https://github.com/qcscine/parrot}) and PyPI (\url{https://pypi.org/project/scine-parrot}).

\section*{Acknowledgment}

This work was supported by an ETH Zurich Postdoctoral Fellowship and by the NCCR Catalysis (grant number 180544), a National Centre of Competence in Research funded by the Swiss National Science Foundation.

\section*{Supporting Information}

Summary of the HDNNP method, description of eeACSF descriptors, tables of eeACSF parameter values, tables of reference element energy values, table of reference data composition (in terms of data origin), evolution of test RMSEs during lMLP training, and performance monitoring of incremental CRN data training without lifelong learning (PDF file).

%
\end{document}